\documentclass[journal,10pt]{IEEEtran}
\usepackage{graphicx,amssymb,lineno}
\usepackage{amsmath,amsfonts,amssymb}
\usepackage{subeqnarray}
\usepackage{cases}
\usepackage{mathrsfs}
\allowdisplaybreaks[4]

\newtheorem{lemma}{Lemma}

\usepackage{algorithm}
\usepackage{algorithmic}
\usepackage[usenames]{color}
\usepackage{float}
\usepackage{graphics}
\usepackage{booktabs}
\usepackage{multicol}
\usepackage{multirow}
\usepackage[noadjust]{cite}
\usepackage[cmyk]{xcolor}
\usepackage{color}

\DeclareMathOperator*{\tr}{tr}

\newcommand\Ab{\ensuremath{{\boldsymbol A}}}
\newcommand{\Phib}{{\boldsymbol{\mathnormal\Phi}}}

\newcommand\Cb{\ensuremath{{\boldsymbol C}}}

\newcommand\Hb{\ensuremath{{\boldsymbol H}}}

\newcommand\Ib{\ensuremath{{\boldsymbol I}}}

\newcommand\Rb{\ensuremath{{\boldsymbol R}}}

\newcommand\Xb{\ensuremath{{\boldsymbol X}}}

\newcommand{\Omegab}{{\boldsymbol{\mathnormal\Omega}}}

\newcommand{\Sigmab}{{\boldsymbol{\mathnormal\Sigma}}}
\newcommand{\Gammab}{{\boldsymbol{\mathnormal\Gamma}}}

\newcommand\phib{\ensuremath{{\boldsymbol \phi}}}

\newcommand\hb{\ensuremath{{\boldsymbol h}}}

\newcommand\kb{\ensuremath{{\boldsymbol k}}}

\newcommand\lb{\ensuremath{{\boldsymbol l}}}

\newcommand\xb{\ensuremath{{\boldsymbol x}}}
\newcommand\yb{\ensuremath{{\boldsymbol y}}}
\newcommand\zb{\ensuremath{{\boldsymbol z}}}
\newcommand\gammab{\ensuremath{{\boldsymbol \gamma}}}

\newcommand\mub{\ensuremath{{\boldsymbol \mu}}}

\newcommand\xib{\ensuremath{{\boldsymbol \xi}}}

\DeclareMathOperator*{\argmax}{argmax}

\makeatletter

\newcommand{\Rmnum}[1]{\expandafter\@slowromancap\romannumeral #1@}
\makeatother

\makeatother

\usepackage{graphicx,graphics,color,epsfig,subfigure,graphpap,rotate}
\usepackage{times, verbatim, subfigure, epsfig, graphicx, latexsym, amsmath}
\usepackage{url}
\usepackage{subfigure}
\usepackage{geometry}
\begin{document}

\title{Off-grid Channel Estimation for Orthogonal Delay-Doppler Division Multiplexing Using Grid Refinement and Adjustment
}
%
\author{Yaru~Shan,~\IEEEmembership{Student Member,~IEEE}, Akram Shafie,~\IEEEmembership{Member,~IEEE}, Jinhong Yuan,~\IEEEmembership{Fellow,~IEEE}, and Fanggang~Wang,~\IEEEmembership{Senior Member,~IEEE}
\thanks{Yaru Shan and Fanggang Wang are with School of Electronic and Information Engineering, Beijing Jiaotong University, Beijing, China (e-mail: yarushan@bjtu.edu.cn, wangfg@bjtu.edu.cn). A. Shafie and J. Yuan are with the School of Electrical Engineering and
Telecommunications, The University of New South Wales, Sydney, NSW, 2052, Australia (e-mail: akram.shafie@unsw.edu.au, j.yuan@unsw.edu.au).
A preliminary version of this work was accepted for presentation at the 2024 IEEE International Conference on Communications Workshop  (ICCW).
}

}

\maketitle

\begin{abstract}
Orthogonal delay-Doppler (DD) division multiplexing (ODDM) has been recently proposed as a promising multicarrier modulation scheme to tackle Doppler spread in high-mobility environments. Accurate channel estimation is of paramount importance to guarantee reliable communication for the ODDM, especially when the delays and Dopplers of the propagation paths are off-grid.
In this paper, we propose a novel grid refinement and adjustment-based sparse Bayesian inference (GRASBI) scheme for DD domain channel estimation.
The GRASBI involves first formulating the channel estimation problem as a sparse signal recovery through the introduction of a virtual DD grid. Then, an iterative process is proposed that involves (i) sparse Bayesian learning  to estimate the channel parameters and (ii) a novel grid refinement and adjustment process to adjust the virtual grid points.
The grid adjustment in GRASBI relies on the maximum likelihood principle to attain the adjustment and utilizes refined grids that have much higher resolution than the virtual grid.
Moreover, a low-complexity grid refinement and adjustment-based channel estimation scheme is proposed, that can provides a good tradeoff between the estimation accuracy and the complexity. Finally, numerical results are provided to demonstrate the accuracy and efficiency of the proposed channel estimation schemes.
\end{abstract}


\section{Introduction}
The next generation wireless systems for 6G and beyond are anticipated to support reliable communication in high-speed scenarios such as the high-speed railway, the low Earth orbit satellite, the unmanned aerial vehicle, etc. \cite{LEO2,UAV2,HSR}.
The severe Doppler spread experienced in these high-speed scenarios undermines the orthogonality among the sub-carriers in the conventional orthogonal
frequency division modulation (OFDM), thereby deteriorating its performance \cite{OFDMICI}.
To guarantee reliable communications and facilitate the diverse emerging applications in high mobility scenarios, an alternative modulation scheme is spurred to propose \cite{OTFSAppear1}.

Delay-Doppler (DD) domain modulations have been proposed to overcome the high Doppler effect in high mobility scenarios \cite{OTFSAppear1}.
In DD modulation, information-bearing symbols are mapped into the DD domain and they are allowed to couple with the doubly selective channel, which has a stable and sparse representation in the DD domain. These enable the DD modulation to achieve full channel diversity \cite{diversity}.
The orthogonal time frequency space (OTFS) modulation is one of the earliest introduced  DD domain modulation scheme \cite{OTFSAppear1}.
However, OTFS encounters two main obstacles during  practical implementation related to the pulse and the out-of-band emission (OOBE) \cite{MP2,SC}.
To address these issues, very recently, a promising alternative for OTFS, known as orthogonal DD division multiplexing (ODDM) modulation, has been proposed \cite{ODDM1,ODDM4}. Specifically, the ODDM signal is generated by aggregating the time and frequency-shifted versions of the DD orthogonal pulse (DDOP) as in a conventional multi-carrier system. The DDOP exhibits orthogonality concerning the delay and Doppler resolutions, and it can satisfy the constraints of the signals in both time and frequency. Thus, the ODDM overcomes the above-mentioned two limitations in the OTFS \cite{SC,qingqing2}.

The DD domain channel estimation plays an important role in ODDM and all the other DD domain modulation schemes \cite{EMB,shencheng}.
The DD domain channel presents quasi-static, path separability, and compactness characteristics. Moreover, the information-bearing symbols in  ODDM   have a direct coupling with the DD domain channel. Due to these, the DD domain input-output relations for ODDM could potentially have a compact representation to benefit and simplify the DD domain channel estimation.
However, the available time duration and the system bandwidth to transmit an ODDM signal are always finite. This results in any arbitrary path delay and Doppler of the physical channel not lying on the exact discretized DD grid in which the information-bearing symbols are. This causes thereby
the phenomenon commonly referred to as  \textit{off-grid delays and Dopplers} \cite{WeiOGSBI,OGCE2,PeakCE,ShanTWC2} or \textit{grid mismatch} \cite{GM1}.
The off-grid delays and  Dopplers result in the  equivalent sampled DD domain channel experiencing spreading effects along the delay and Doppler dimensions, respectively, which deteriorate its sparsity \cite{Akram2,ODDMIO}. This poses diverse challenges for channel estimation.

\enlargethispage{0.20 cm}
Several studies have investigated DD domain channel estimation \cite{EMB,COR,Akram,WeiOGSBI,OGCE2,Shan2,ShanTWC2}.
For instance, in \cite{EMB},  a threshold-based approach was proposed to estimate the equivalent sampled DD domain channel by detecting the amplitude of the received signal. However,  \cite{EMB} proposed embedding guard symbols
to prevent the pilot from being polluted by the surrounding data symbols.
Due to this, when considering the off-grid delays and off-grid Dopplers, considerable symbols in the transmission frame have to be set as guard symbols, thereby reducing the spectral efficiency.
In \cite{COR}, the off-grid Doppler was estimated by thresholding the cross-correlation function between the pilot impulse and the Doppler function.
However, to obtain the channel response, only the pilot symbol is transmitted in one frame, which also dramatically reduces the spectral efficiency. Moreover, in \cite{Akram}, an interference cancellation-based DD domain channel estimation technique was proposed for OTFS when it coexists with an OFDM communication system.
However,  \cite{Akram} assumed the delays and the Dopplers of the channel are on-grid. When extending the algorithms in  \cite{Akram} to the off-grid  condition, the channel estimation performance degrades \cite{mismatch}.

To improve the off-grid channel estimation accuracy, the compressed sensing principles \cite{CS2,CS3} have been recently studied in \cite{WeiOGSBI,OGCE2}. To take advantage of the sparsity of the DD domain channel response, they aim to estimate the DD domain channel response rather than the equivalent sampled DD domain channel \cite{CS2,CS3}. In particular, \cite{WeiOGSBI} first transformed the off-grid channel estimation problem into the sparse recovery by introducing a virtual DD grid \cite{WeiOGSBI}. Different from the DD grid in which  the information-bearing symbols are mapped, the virtual DD grid has higher delay and the Doppler resolutions.
Thereafter, the on-grid and off-grid elements of channel taps with respect to the virtual grid were introduced. Then, using the principle of sparse Bayesian learning (SBL), an off-grid sparse Bayesian inference (OGSBI) scheme  was proposed. As the SBL is guided by the maximum likelihood (ML) principle, the OGSBI was able to achieve high off-grid  channel estimation accuracy.

Despite the benefits of OGSBI, it introduced a linear approximation error while separating the estimation for the on-grid and off-grid elements (see (22) in \cite{WeiOGSBI}).
To overcome this limitation, a grid evolution process was introduced for DD domain channel estimation in \cite{Shan2}. The grid evolution-based sparse Bayesian inference algorithm proposed in \cite{Shan2} was referred to as GESBI. In GESBI, after the SBL, the virtual DD grid locations nearest to the channel taps were  adjusted to ensure that the off-grid elements of channel taps reach zero. This grid evolution process certainly mitigates the approximation error in \cite{WeiOGSBI}, and thus GESBI outperforms OGSBI.
The OGSBI encounters  two drawbacks. First, its grid evolution process directly relies on the delay and the Doppler estimates from the SBL. This limits the improvements of the channel estimation that happen through grid evolution. Second, the computational complexity of GESBI was high due to the introduction of the grid evolution process.
To reduce the complexity of GESBI, grid evolution-based efficient sparse Bayesian inference with Student's t distribution referred to as
T-GEESBI was proposed in \cite{ShanTWC2}. The T-GEESBI relied on efficient SBL to reduce the complexity. Despite its complexity benefits, its accuracy performance degraded compared to the GESBI. Moreover, as in GESBI, the grid evolution process in T-GEESBI directly relied on the delay and the Doppler estimates from the SBL. The drawbacks of OGSBI, GESBI, and T-GEESBI motivated this work to development of a novel channel estimation scheme.

Very recently, \cite{SBL2} proposed a novel one-dimensional gridless DoA estimation algorithm using a novel grid refinement process. Taking inspirations from \cite{SBL2}, in this work, we propose novel channel estimation schemes which rely on \textit{grid refinement and adjustment}, instead of grid evolution as in GESBI and T-GEESBI.
The main contributions of this work are summarized in the following:

\textbf{1)} To address the off-grid channel estimation in the ODDM system, we  proposed a grid refinement and adjustment-based sparse Bayesian inference channel estimation scheme, which we refer to as GRASBI. Our proposed GRASBI first performs the SBL to estimate the channel parameters. Different from OGSBI,  GESBI, and  T-GEESBI, our algorithm does only introduce on-grid elements of channel taps with respect to the virtual grid, while aiming to tackle the potential off-grid elements of channel taps with respect to the virtual grid, through grid refinement and adjustment. In doing so, the proposed GRASBI avoids the linear approximation error experienced in the above-mentioned channel estimation schemes.

\textbf{2)} We then propose a grid refinement and adjustment in GRASBI to adjust the virtual grid locations nearest to the channel taps. In particular, first refined grids, which have much higher resolution as compared to the virtual grid, are generated around each peak. Then,  the  virtual grid locations nearest to the channel taps are sequentially adjusted such that they now lie on the refined grid. The grid adjustment is carried out to minimize the estimation error according to the ML principle while simultaneously updating  the channel estimation parameters.

\textbf{3)} Considering the computational complexity of the proposed GRASBI, we propose a low-complexity  grid refinement and adjustment efficient sparse Bayesian inference with Student's t distribution scheme referred to as T-GRAESBI. In T-GRAESBI,  the efficient sparse Bayesian inference is utilized to transform the large-scale matrix inversion problem into the diagonal matrix inversion and the  \textit{Student's t} distribution is used as the prior distribution.
These lead to significant reduction in the computation complexity.

\textbf{4)} We numerically evaluate the estimation accuracy and the efficiency of our proposed GRASBI and T-GRAESBI channel estimation schemes and arrive at the following conclusions.
\begin{itemize}
\item The proposed GRASBI can outperform the SBL-based DD domain channel estimation schemes for ODDM, such as OGSBI  \cite{WeiOGSBI}, GESBI \cite{Shan2}, and T-GEESBI \cite{ShanTWC2}.
\item While significantly reducing the computational complexity, the proposed T-GRAESBI achieves a  channel estimation accuracy that is slightly lower than that of proposed GRASBI, but higher than those of OGSBI, GESBI and T-GEESBI. Thus, T-GRAESBI provides a good tradeoff between the accuracy and the complexity for ODDM systems.
\item The grid refinement and adjustment in our proposed GRASBI and T-GRAESBI are robust under different channel conditions. Moreover, the channel estimation accuracy improves when increasing the number of the exterior iteration of the grid refinement and adjustment.
\end{itemize}

The remainder of the paper is organized as follows. Section II introduces the ODDM system model. In Section III, the proposed grid refinement and adjustment scheme GRASBI is illustrated, which is followed by the introduction of the proposed low-complexity T-GRAESBI scheme in Section IV.
Then, Section V provides the numerical results. Finally, we conclude this paper in Section VI.
\begin{figure*}[tbp]
\centering
\centerline{\includegraphics[width=0.95\textwidth]{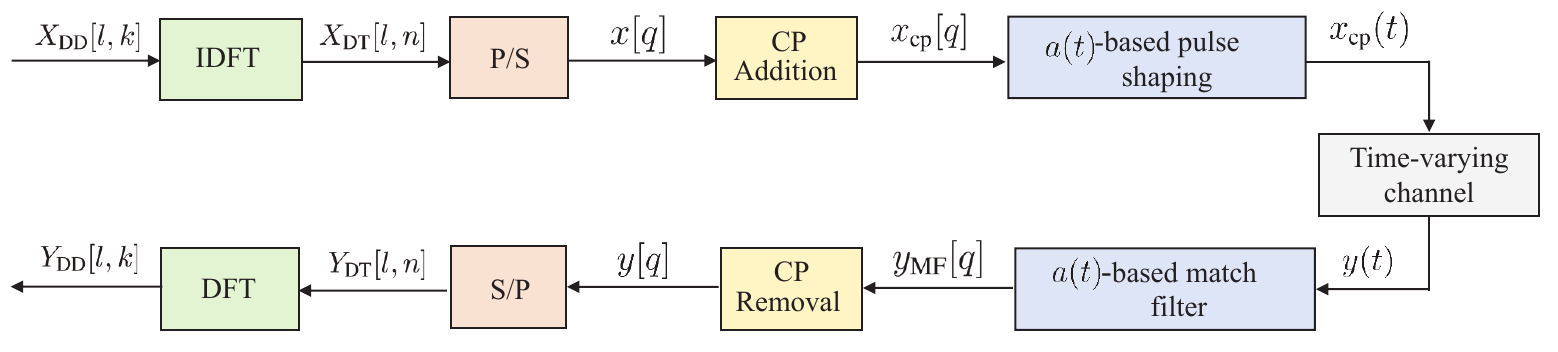}}
\caption{Illustration of the ODDM system.}
\label{fig}
\end{figure*}

\emph{Notation}: Boldface lowercase
and uppercase letters denote vectors and matrices, respectively. The operators $(\cdot)^{\ast}$, $(\cdot)^\mathrm{T}$, $(\cdot)^\mathrm{H}$, $[\cdot]_{M}$,  $(\cdot)^{-1}$, $\tr\{\cdot\}$, and $\text{eig}(\cdot)$ denote the conjugate,  the transpose, the Hermitian, the modulo $M$, the inverse, the trace, and the maximum eigenvalue of their arguments, respectively.
For any vector $\xb$, $\text{diag}\{\xb\}$ denotes a diagonal square matrix whose diagonal is that of $\xb$. $\Ib_{M}$ denotes the identity matrix of size $M\times M$.  $\mathcal{I}_{\varpi}\triangleq[0,1,\dots,\varpi-1]$  denotes the shorthand to represent an index set, for all $\varpi \in \{M,N, MN,\dots\}$.
A complex random vector $\xb$ which follows a complex Gaussian distribution with mean vector $\mub$ and covariance matrix $\Sigmab$ is defined as $\mathcal{CN}(\xb;\mub,\Sigmab)=\frac{1}{\pi^{N}|\Sigmab|}e^{-(\xb-\mub)^{\text{H}}\Sigmab^{-1}(\xb-\mub)}$. A random variable $x$ which follows  a Gamma distribution is defined as $\Gamma(x;a,b)=\frac{b^{a}x^{a-1}e^{-bx}}{\Gamma(a)}$ with $a>0$, $b>0$, and $\Gamma(a)$  being Gamma function.

\section{System Model}\label{formulation}
In this section,  the ODDM system model and it's DD domain input-output relation are detailed.
\subsection{ODDM System Model}
The transmission and reception processes for the ODDM system are described in Fig. $1$.
\subsubsection{Transmitter}
The ODDM modulation scheme was proposed in \cite{ODDM1}. At the transmitter of an ODDM system, the $MN$ information-bearing symbols  are first mapped into the DD domain to obtain $X_{\text{DD}}[l,k]$, where $l\in\mathcal{I}_{M}$ and $k\in\mathcal{I}_{N}$ are the delay index and the Doppler index, respectively, and $M$ and $N$ are the numbers of the ODDM symbols and the subcarriers, respectively. The ODDM system has bandwidth $\frac{M}{T}$ and frame duration $NT$. Each multicarrier signal's subcarrier spacing is $\frac{1}{NT}$ and $T$ is multi-carrier symbol interval.

Then the discrete delay-time (DT) domain signal is generated by performing the inverse discrete Fourier transform (IDFT) on $\Xb_{\text{DD}}$ as
\begin{align}
X_{\text{DT}}[l,n]=& \frac{1}{\sqrt{N}}\sum_{k=0}^{N-1}X_{\text{DD}}[l,k]e^{j2\pi\frac{nk}{N}},\quad  n\in\mathcal{I}_{N}, l\in\mathcal{I}_{M}.
\end{align}
Thereafter, by serializing $X_{\text{DT}}[l,n]$, the discrete time domain signal is obtained as $x[q]=X_{\text{DT}}[[q]_{M},\lfloor\frac{q}{M}\rfloor]$, where $q\in\mathcal{I}_{MN}$ is the discrete time index. Next, considering the maximum delay spread of the channel $\tau_{\text{max}}$, a cyclic prefix (CP) with length $T_{\text{cp}}=D\frac{T}{M}>\tau_{\text{max}}$, $D\in\mathbb{Z}$, is added to mitigate the inter frame interference, where $D$ is the number of samples for the CP and $T$ is the symbol interval. In doing so, the time domain transmitted signal $x_{\text{cp}}[q]$ is obtained.
Finally, the  sample-wise pulse shaping is performed on $x_{\text{cp}}[q]$ to obtain the transmit-ready time domain signal as
\begin{align}
x_{\text{cp}}(t)=\sum_{q=-N_{\text{cp}}}^{MN-1}x_{\text{cp}}[q]a\left(t-q\frac{T}{M}\right),
\end{align}
where $a(t)$ is the square root Nyquist pulse with the sampling period $T_{\text{s}}=\frac{T}{M}$ and the duration $KT_{\text{s}}$, where $K\ll M$.\footnote{The ODDM was originally proposed to directly modulate $X_{\text{DD}}[l,k]$ using the DDOP $u(t)=\sum_{\dot{n}=0}^{N-1}a(t-\dot{n}T)$, which satisfies the DD domain bi-orthogonality condition \cite{ODDM1}. However, the $a(t)$-based implementation of ODDM provides a close approximation for the $u(t)$-based implementation of ODDM, but with a relatively simple implementation process \cite{ODDM1}. Considering this, in this work, we adopt the $a(t)$-based implementation at the transmitter/receiver for ODDM.}

\subsubsection{Channel} For doubly-selective channels, the DD domain spread function $h(\tau,\nu)$ is expressed  in \cite{MP2}
\begin{align}
h(\tau,\nu)=\sum_{p=1}^{P}\rho_{p}\delta(\tau-\tau_{p})\delta(\nu-\nu_{p}), \label{eq:3}
\end{align}
where $\rho_{p}\in\mathbb{C}$, $\tau_{p}\in(0,\tau_{\text{max}})$, and  $\nu_{p}\in(-\nu_{\text{max}},\nu_{\text{max}})$ are the channel coefficient, the delay, and the Doppler of the $p$th propagation path, respectively, $\nu_{\text{max}}$ is the maximum Doppler of the channel, and $P$ is the total number of propagation paths in the channel. When considering the delay and Doppler resolutions of ODDM, $\frac{T}{M} $ and $\frac{1}{NT}$, respectively, (3) can be rewritten as
\begin{align}
h(\tau,\nu)=\sum_{p=1}^{P}\rho_{p}\delta\left(\tau-\frac{l_{p}T}{M} \right)\delta\left(\nu-\frac{k_{p}}{NT}\right), \label{eq:3}
\end{align}
where $l_{p}=\frac{\tau_{p}}{T_{\text{s}}}\in\mathbb{R}$ and $k_{p}=\nu_{p}NT\in\mathbb{R}$ are the normalized delay and  Doppler of the $p$th path, respectively. We note that as $\tau_{p}\in\mathbb{R}$, $\nu_{p}\in\mathbb{R}$, and the fact that system bandwidth and duration are always finite,
the normalized delays $l_{p}$ and  Dopplers $k_{p}$, may not always fall onto the exact DD grid defined by the delay and Doppler resolutions of ODDM. This leads to $l_{p}\in\mathbb{R}$ and $k_{p}\in\mathbb{R}$, which refer to the  off-grid delays and off-grid Dopplers phenomenon  \cite{WeiOGSBI}.
\enlargethispage{0.15 cm}

After passing through the doubly-selective channel characterized by  \eqref{eq:3},  the time-domain received signal becomes
\begin{subequations}
\begin{align}
y(t)=&\int_{-\nu_{\text{max}}}^{\nu_{\text{max}}}\int_{0}^{\tau_{\text{max}}} h(\tau,\nu)e^{j2\pi\nu(t-\tau)}x_{\text{cp}}(t-\tau){\rm d}\tau{\rm d}\nu+z(t) \\
=& \sum_{p=1}^{P}\rho_{p}x_{\text{cp}}\left(t-\frac{l_{p}T}{M}\right)e^{j2\pi\frac{k_{p}}{NT}\left(t-\frac{l_{p}T}{M}\right)}+z(t),
\end{align}
\end{subequations}
where $z(t)\sim\mathcal{CN}(0, \sigma^{2})$ denotes the time-domain circularly symmetric complex Gaussian  noise.

\subsubsection{Receiver}
At the receiver, the $a(t)$-based matched filtering  is first performed on $y(t)$ to obtain
\begin{align}
y_{\text{MF}}(t)=\int_{\tau'}y(t-\tau')a^{\ast}(-\tau'){\rm d}\tau'.
\end{align}
Next,  $y_{\text{MF}}(t)$ is sampled at $t=qT_{\text{s}}$ yielding $y_{\text{MF}}[q]$. When removing the CP and considering $d=q-q'$ as the lag between the transceiver, the received sequence $y[q]$ can be expressed as
\begin{align}
y[q]=\sum_{d=0}^{D-1}h[d,q]x[q-d]+z[q],
\end{align}
where  $q$ is the discrete time index and
\begin{align}
h[d,q]\triangleq\sum_{p=1}^{P}\rho_{d,p}e^{j2\pi\frac{k_{p}}{NT}\left(qT_{\text{s}}-\frac{l_{p}T}{M}\right)}, \ d\in\mathcal{I}_{D},\ q\in\mathcal{I}_{MN},
\end{align}
$\rho_{d,p}\triangleq\rho_{p}g(dT_{\text{s}}-\tau_{p})$, and $g(\tau)\triangleq a(\tau)\ast a^{\ast}(-\tau)$ denotes the effective pulse, which characterizes the overall effect of the transceiver filters \cite{ODDMIO}. After parallelizing $y[q]$, the DFT is performed to obtain the DD domain received signal  as
\begin{align}
Y_{\text{DD}}[l,k]=\frac{1}{\sqrt{N}}\sum_{n=0}^{N-1}Y_{\text{DT}}[l,n]e^{-j2\pi\frac{nk}{N}},
\end{align}
where $Y_{\text{DT}}[l,n]=y[nM+l]$.
\subsection{ODDM Input-output Relation}
For the ODDM system described above, the DD domain input-output relation can be derived as \cite{ODDMIO}
\begin{align}
Y_{\text{DD}}[l,k]=& \sum_{p=1}^{P}\sum_{d=0}^{D-1}\sum_{\tilde{n}=0}^{N-1}h[l,d,p][\Ab_{\nu,p}]_{k,\tilde{n}}\Psi[l,d,\tilde{n}]\nonumber\\
&\qquad\times X_{\text{DD}}[[l-d]_{M},\tilde{n}]+Z_{\text{DD}}[l,k],\label{eq:11}
\end{align}
where
\begin{align}
 h[l,d,p]&\triangleq\rho_{p}g((d-l_{p})T_{s})e^{j2\pi\frac{(l-l_{p})k_{p}}{MN}}, \ l\in\mathcal{I}_{M},\ d\in\mathcal{I}_{D},\\
 [\Ab_{\nu,p}]_{k,\tilde{n}}&\triangleq\frac{1}{N}\frac{1-e^{j2\pi(\tilde{n}+k_{p}-k)}}{1-e^{j\frac{2\pi}{N}(\tilde{n}+k_{p}-k)}},\ k,\tilde{n} \in\mathcal{I}_{N},
\end{align}
\begin{subequations}
\begin{numcases}{\Psi[l,d,\tilde{n}]=}
1,  \quad\quad\quad\  {0\leq d\leq l\leq M-1,}     \\
e^{-j2\pi\frac{\tilde{n}}{N}},    \ \ \ {0\leq l<d \leq D-1.}
\end{numcases}
\end{subequations}
When $Y_{\text{DD}}[l,k]$, $X_{\text{DD}}[l,k]$, and $Z_{\text{DD}}[l,k]$ in \eqref{eq:11} are vectorized along the Doppler dimension to form $\yb_{\text{DD}}$, $\xb_{\text{DD}}$, and $\zb_{\text{DD}}$, respectively, the DD domain input-output relation for ODDM can be given in the matrix-vector form as $\yb_{\text{DD}}=\Hb_{\text{DD}}\xb_{\text{DD}}+\zb_{\text{DD}}$, where $\Hb_{\text{DD}}\in\mathbb{C}^{MN\times MN}$ is the equivalent sampled DD domain channel matrix.

\section{Grid Refinement and Adjustment-based Channel Estimation for ODDM Systems}
In this section, a novel channel estimation scheme for ODDM  is introduced.  First, the
details of the sparse signal recovery problem, which reflect the channel estimation problem, are provided. Then, the proposed grid refinement and adjustment-based channel estimation scheme, referred to as GRASBI, is introduced.
\begin{figure}[tbp]
\centering
\centerline{\includegraphics[width=0.50\textwidth]{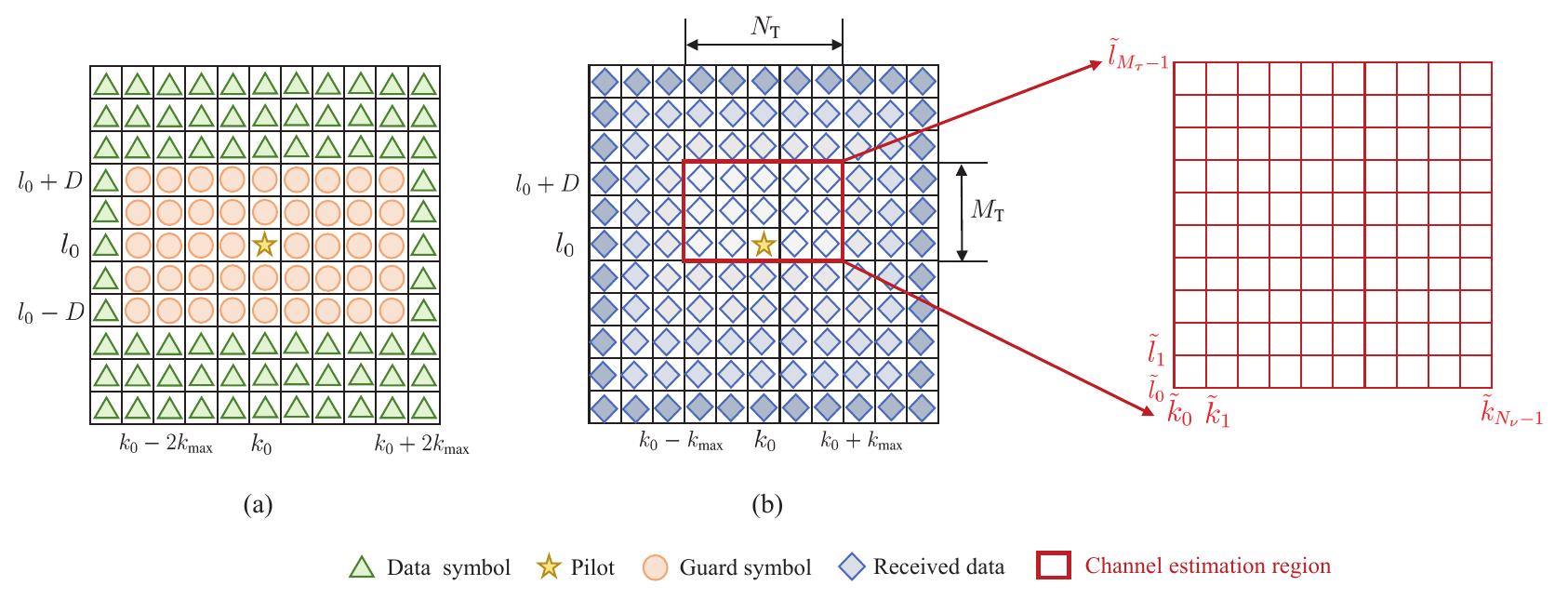}}
\caption{Illustration of (a) the transmitted DD domain signal/frame and (b) the received DD domain signal/frame.
 }
\label{fig}
\end{figure}
\subsection{Problem Description}
\subsubsection{Pilot Pattern and Channel Estimation Region} To facilitate channel estimation, we first rearrange $X[l,k]$ by adding a pilot  and  guard symbols surrounding the pilot symbol in a manner that mitigates the interference between the pilot and data symbols in the DD domain \cite{EMB}. In particular, after including the pilot and guard symbols, $X_{\text{DD}}[l,k]$ becomes\footnote{The designed pilot pattern can be directly extended to multiple pilots. Here, we only set one pilot to simplify the illustration.}
\begin{subequations}
\begin{numcases}{X_{\text{DD}}[l,k]=}
d_{0},  \quad\quad  {l=l_{0},\ k=k_{0}, }     \\
0,      \quad\quad\  {l\in [l_{0}-D, l_{0}+D],}\\
\quad   \quad\quad\  {k\in [k_{0}-2k_{\text{max}},k_{0}+2k_{\text{max}}],}\nonumber\\
d[l,k], \,\ {\text{otherwise},}
\end{numcases}
\end{subequations}
where $d_{0}$ and $d[l,k]$ denote the pilot and the data symbols, respectively. Due to the doubly selective nature of the channel and the fact that delays and Dopplers are off-grid, the pilot energy will appear at all indices in $Y_{\text{DD}}[l,k]$  \cite{WeiOGSBI,Shan2}. Despite this, similar to that in \cite{WeiOGSBI},  we will utilize the received pilot signal corresponding to a certain region of the $Y_{\text{DD}}[l,k]$ where the pilot energy will be predominantly concentrated. We refer to this region as the \textit{channel estimation region}, and it corresponds to the indices, as shown in Fig. $2$(b), $l\in [l_{0}, l_{0}+D]$ along the delay dimension and $k\in [k_{0}-k_{\text{max}},k_{0}+k_{\text{max}}]$ along the Doppler dimension.

\begin{figure*}[tbp]
\centering
\centerline{\includegraphics[width=1.00\textwidth]{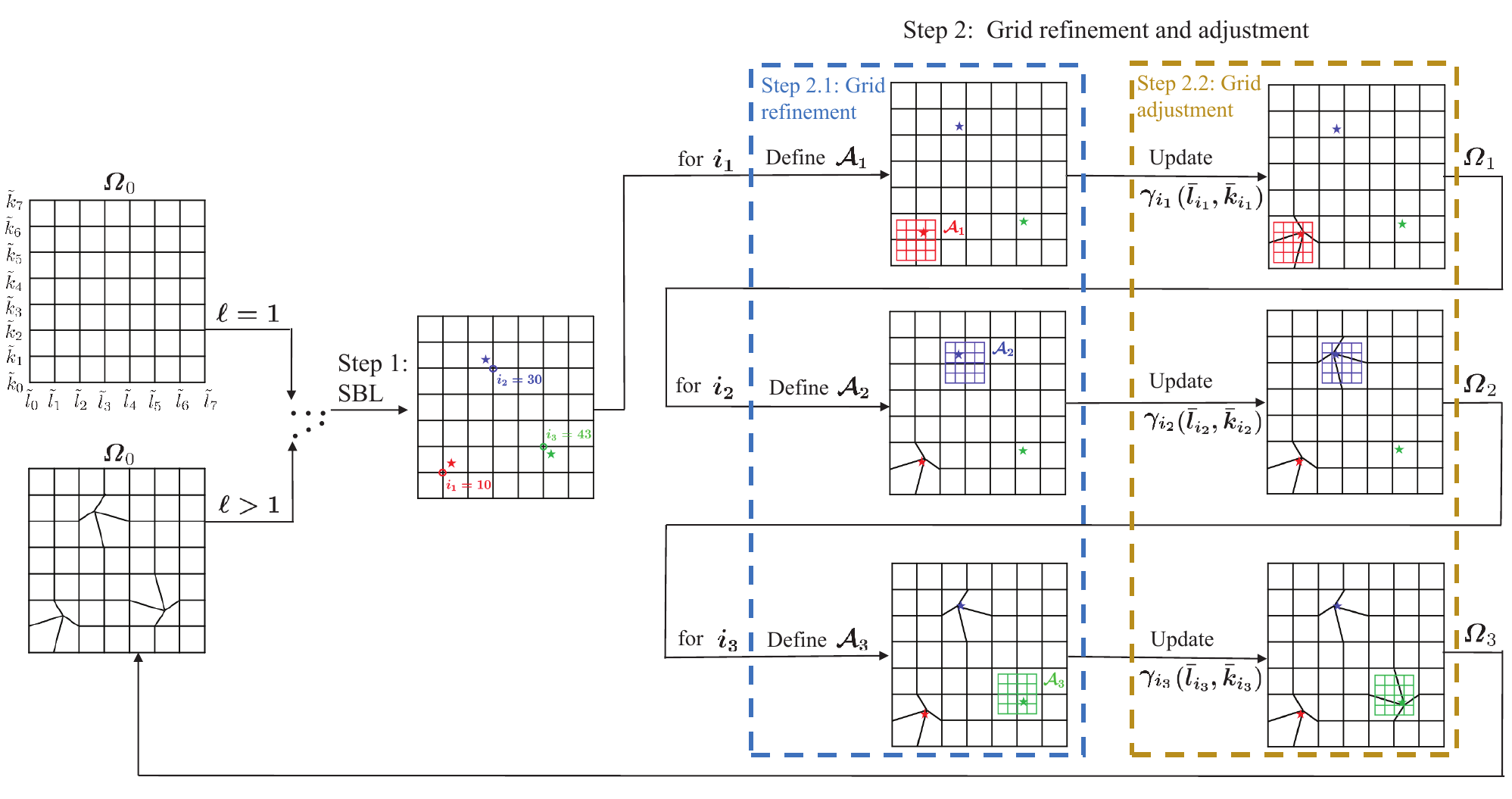}}
\caption{Illustration of the proposed GRASBI for one iteration. The three stars in the figure indicate the position of the three paths of the channel. The delay and the Doppler indices of the channel  to be estimated are $[0.31, 0.78, 1.37]$ and $[-1.25, 0.93,-0.95]$, respectively. The size of the channel estimation region is set as  $M_{\text{T}}=D+1=3$ and $N_{\text{T}}=2k_{\text{max}}+1=5$. The size of the uniform virtual grid $\Omegab_{0}$ is set as $M_{\tau}=N_{\nu}=8$.
 }
\label{fig}
\end{figure*}
Using \eqref{eq:11}, $Y_{\text{DD}}[l,k]$ in the channel estimation region can be expressed as (15), which is shown at the top of the next page.
In (15), the $\tilde{Z}[l,k]$ is the measurement noise\footnote{We clarify that at the receiver,  $\tilde{Z}[l, k]$  is not measured. Differently, only the received signal $Y_{\text{DD}}[l, k]$ is measured. However, following the existing literature \cite{WeiOGSBI,OGCE2}, we refer to the term $\tilde{Z}[l,k]$ as the measurement noise.} which includes not only the noise but also the interference from the data symbols $d[l,k]$.
\newcounter{tempeqncnt}
\setcounter{tempeqncnt}{\value{equation}}
\setcounter{equation}{14}
\begin{figure*}[!t]
\normalsize
\begin{align}
Y_{\text{DD}}[l,k]=& \sum_{p=1}^{P}h[l,l-l_{0},p][\Ab_{\nu,p}]_{k,k_{0}}d_{0}+\underbrace{\sum_{p=1}^{P}\sum_{d=0,\atop d\neq l-l_{0}}^{D-1}\sum_{\tilde{n}=0, \atop\tilde{n}\neq k_{0}}^{N-1}h[l,d,p][\Ab_{\nu,p}]_{k,\tilde{n}}\Psi[l,d,\tilde{n}]X_{\text{DD}}[[l-d]_{M},\tilde{n}]+Z_{\text{DD}}[l,k]}_{\tilde{Z}[l,k]}.
\label{eq:172}
\end{align}
\setcounter{tempeqncnt}{\value{equation}}
\hrulefill
\vspace*{3pt}
\end{figure*}
Based on (15), the vectorized input-output relation for ODDM  in the channel estimation region can be obtained as
\begin{align}
\yb_{\text{T}}=\Phib\hb+\tilde{\zb}, \label{eq:18}
\end{align}
where $\hb=[\tilde{\rho}_{1},\tilde{\rho}_{2},\dots,\tilde{\rho}_{P}]^{\text{T}}\in\mathbb{C}^{P}$, $\tilde{\rho}_{p}=\rho_{p} e^{-j2\pi\frac{l_{p}k_{p}}{MN}}$, $\yb_{\text{T}}\in\mathbb{C}^{M_{\text{T}}N_{\text{T}}}$, $M_{\text{T}}=D+1$, $N_{\text{T}}=2k_{\text{max}}+1$, $\Phib=\big[\phib(l_{1},k_{1}),\phib(l_{2},k_{2}),  \dots,\phib(l_{P},k_{P})\big]\in\mathbb{C}^{M_{\text{T}}N_{\text{T}}\times P}$ is the measurement matrix, and $\phib(l_{p},k_{p})\in\mathbb{C}^{M_{\text{T}}N_{\text{T}}}$ is the $p$th column vector of $\Phib$ given by
\begin{align}
\phib(l_{p},k_{p})=&~ g((l-l_{0}-l_{p})T_{s})e^{j2\pi\frac{lk_{p}}{MN}}
\frac{d_{0}}{N}\frac{1-e^{j2\pi(k_{0}+k_{p}-k)}}{1-e^{j\frac{2\pi}{N}(k_{0}+k_{p}-k)}}.
\end{align}

\subsubsection{Channel Estimation as a Sparse Signal Recovery Problem}  Note that although the values of $\yb_{\text{T}}$ in \eqref{eq:18} is readily available at the receiver, channel parameters cannot be directly estimated  using \eqref{eq:18}. This is because of two reasons. First, both $\Phib$ and $\hb$  are not known since they are characterized based on the unknown $l_{p}$, $k_{p}$, and $h_{p}$. Second, the dimension of $\Phib$ and $\hb$ are also unknowns, since the total number of channel taps $P$, which determines the size of the $\Phib$ and the $\hb$, is not known.
Considering these challenges,  we  first transform \eqref{eq:18} into a sparse recovery problem, the details are given next.

As shown in Fig. $2$(b), we first partition the channel estimation region using the  virtual DD grid  $\Omegab_0$ with the size $M_{\tau}\times N_{\nu}$.
Note that the suitable values for $M_{\tau}$ and $N_{\nu}$ need to be determined, which will be discussed later in this subsection.
We clarify that the virtual DD grid is different from the DD grid with resolutions $\frac{T}{M}$ and $\frac{1}{NT}$, which is used to modulate the data symbols.
Under the uniformly distributed, the virtual delay and the virtual Doppler resolutions become $r_{\tau}=\frac{D}{M_{\tau}-1}$, and $r_{\nu}=\frac{2k_{\text{max}}}{N_{\nu}-1}$, respectively.
We then define the virtual delay vector $\bar{\lb}=\big[\bar{l}_{0},\bar{l}_{1},\dots,\bar{l}_{M_{\tau}N_{\nu}-1}\big]$ and the virtual Doppler vector $\bar{\kb}=\big[\bar{k}_{0},\bar{k}_{1},\dots,\bar{k}_{M_{\tau}N_{\nu}-1}\big]$. For any virtual grid $(\tilde{k}_{a}$, $\tilde{l}_{b})$, $a\in\mathcal{I}_{N_{\nu}}, b\in\mathcal{I}_{M_{\tau}}$,  we have a corresponding $(\bar{k}_{i}$, $\bar{l}_{i})$, $i\in\mathcal{I}_{M_{\tau}N_{\nu}}$, $\bar{k}_{i}=\tilde{k}_{a}$, $\bar{l}_{i}=\tilde{l}_{b}$,  $a=i-bN_{\nu}$, $b=\big\lfloor\frac{i-0.5}{N_{\nu}}\big\rfloor$, $\tilde{k}_{a}=a r_{\nu}-k_{\text{max}}$, $\tilde{l}_{b}=br_{\tau}$. With these definitions, \eqref{eq:18} can be transformed into
\begin{align}
\yb_{\text{T}}=\bar{\Phib}\left(\bar{\lb},\bar{\kb}\right)\bar{\hb}+\tilde{\zb},\label{eq:192}
\end{align}
where $\bar{\hb}\in\mathbb{C}^{M_{\tau}N_{\nu}}$ and $\bar{\Phib}(\bar{\lb},\bar{\kb})\in\mathbb{C}^{M_{\text{T}}N_{\text{T}}\times M_{\tau}N_{\nu}}$ is the new measurement matrix with its $i$th column $\phib(\bar{l}_{i},\bar{k}_{i})$, $i\in\mathcal{I}_{M_{\tau}N_{\nu}}$.

We clarify that after this transformation, unlike \eqref{eq:18}, now the dimension of $\bar{\Phib}\left(\bar{\lb},\bar{\kb}\right)$ and $\bar{\hb}$ in (18) are known.  Moreover, $\bar{\Phib}$ can be determined based on the delay and the Doppler vectors of the virtual DD gird. Furthermore, after the determination of $\bar{\Phib}$, $\bar{\hb}$ can be determined based on (18). We further note that although $\bar{\hb}$ has $M_{\tau}N_{\nu}$ elements, only $P$  elements in it are non-zero if the channel has $P$ separable propagation paths.
Due to this and the fact that $M_{\tau}N_{\nu}\gg P$, $\bar{\hb}$ in (18) exhibits a sparse structure. This enables the development of a sparse recovery problem based on \eqref{eq:192}.  We clarify that despite the channel has $P$ separable propagation paths, leading to $P$ non-zeros in $\bar{\hb}$, it is estimated by $\hat{P}=\big\lfloor\frac{M_{\text{T}}N_{\text{T}}}{\ln(M_{\tau}N_{\nu})}\big\rfloor$ according to compressed sensing  \cite{WeiOGSBI}.

\subsection{Grid Refinement and Adjustment-based Channel Estimation Scheme}
To solve the sparse recovery problem, we propose a grid refinement and adjustment-based iterative
channel estimation scheme  referred to as GRASBI. A simplified visual illustration of our proposed GRASBI scheme for a channel with three paths is presented in Fig. $3$. In each iteration of this scheme, two steps are carried out. In the first step,
the SBL is performed to estimate the channel parameters for a given virtual grid. In the second step, based on the channel parameters estimated in the first step, virtual grid points are sequentially refined and adjusted while simultaneously updating the channel parameters.

\textbf{\emph{Step 1: Sparse Bayesian Learning.}}
The SBL solves sparse signal recovery problems by using the ML  principle; thus, it demonstrates superior accuracy of the sparse recovery \cite{SBL}. To determine $\bar{\hb}$ in \eqref{eq:192}  using the SBL,  $\bar{\hb}$ and the measurement noise $\tilde{\zb}$ are assumed to follow Gaussian distribution as
\begin{align}
P(\bar{\hb}|\Gammab)=& ~\mathcal{CN}(\bar{\hb};\mathbf {0}, \Gammab), \label{eq:22V2}\\
P(\tilde{\zb}|\lambda)=&~ \mathcal{CN}\left(\tilde{\zb};\mathbf {0},\lambda\Ib_{M_{\text{T}}N_{\text{T}}}\right), \label{eq:23V2}
\end{align}
where $\Gammab=\text{diag}\{\gammab\}$, $\gammab=[\gamma_{0},\gamma_{1},\dots,\gamma_{M_{\tau}N_{\nu}-1}]^{\text{T}}$, $\gamma_{i}\geq0$, $i\in\mathcal{I}_{M_{\tau}N_{\nu}}$, and $\lambda$ is the noise variance. The vector $\gammab$ is utilized to control the supports of the channel estimate as $\bar{h}_{i}=0$ when $\gamma_{i}=0$ and $\bar{h}_{i}\neq0$ when $\gamma_{i}>0$.

Then based on \eqref{eq:22V2} and \eqref{eq:23V2}, the likelihood function for $\yb_{\text{T}}$ in \eqref{eq:192} is obtained as
\begin{align}
P(\yb_{\text{T}}|\bar{\hb};\Gammab,\lambda)=\mathcal{CN}\left(0,\bar{\Phib}(\bar{\lb},\bar{\kb})\Gammab\bar{\Phib}^{\text{H}}(\bar{\lb},\bar{\kb})+\lambda\Ib_{M_{\text{T}}N_{\text{T}}}\right). \label{eq:24V2}
\end{align}
Next, the expectation maximization (EM) algorithm proposed in \cite{SBL11} is adopted to update the parameters $\gammab$ and $\lambda$.
In Step E of the EM algorithm, given \eqref{eq:24V2}, $\lambda$, and $\Gammab$, the conditional posterior distribution for $\bar{\hb}$ is obtained as $P(\bar{\hb}|\yb_{\text{T}};\Gammab,\lambda)=\mathcal{CN}(\mub_{\bar{\hb}},\Sigmab_{\bar{\hb}})$, where
\begin{align}
\Sigmab_{\bar{\hb}}=& ~ \Gammab-\Gammab\bar{\Phib}^{\text{H}}\left(\bar{\lb},\bar{\kb}\right)\left(\lambda\Ib_{M_{\text{T}}N_{\text{T}}}+\bar{\Phib}\left(\bar{\lb},\bar{\kb}\right)\Gammab\bar{\Phib}^{\text{H}}\left(\bar{\lb},\bar{\kb}\right)\right)^{-1}\nonumber\\
&\qquad\times\bar{\Phib}\left(\bar{\lb},\bar{\kb}\right)\Gammab, \label{eq:212}\\
\mub_{\bar{\hb}}=& \lambda^{-1}\Sigmab_{\bar{\hb}}\bar{\Phib}^{\text{H}}\left(\bar{\lb},\bar{\kb}\right)\yb_{\text{T}}\label{eq:222}.
\end{align}
In Step M of the EM algorithm, the joint distribution is averaged through the conditional posterior distribution obtained  in Step E. Based on these, the updating rules for $\gammab$ and $\lambda$ are obtained as \cite{SBL11}
\begin{align}
\gamma_{i}=& ~\frac{\|\mub_{\bar{\hb}} \|^{2}}{1-\gamma^{-1}_{i}\Sigmab_{\bar{\hb}_{i,i}}},\quad  i\in\mathcal{I}_{M_{\tau}N_{\nu}},\label{eq:232}\\
\lambda=& ~\frac{\big\| \yb_{\text{T}}-\bar{\Phib}\left(\bar{\lb},\bar{\kb}\right)\mub_{\bar{\hb}}\big\|^{2}}{M_{\text{T}}N_{\text{T}}-M_{\tau}N_{\nu}+\sum_{i=0}^{M_{\tau}N_{\nu}-1}\frac{\Sigmab_{\bar{\hb}_{i,i}}}{\gamma_{i}}}\label{eq:242}.
\end{align}
After the SBL, the channel coefficients can be obtained from the non-zero elements of the mean vector $\mub_{\bar{\hb}}$.
The delay and the Doppler can be estimated based on the virtual grid indices corresponding to the non-zeros of $\mub_{\bar{\hb}}$. A brief summary of
the steps in SBL are given in lines $3$ to $8$ in Algorithm $1$.

\underline{Challenge of only using SBL:} As highlighted in Section II-B-2), the delays and Dopplers of the channel paths are off-grid, i.e., $l_{p}\in\mathbb{R}$ and $k_{p}\in\mathbb{R}$. Due to this, estimating $l_{p}$ and $k_{p}$ as in the above-mentioned SBL based on the indices of the virtual grid leads to the estimation accuracy of $l_{p}$ and $k_{p}$ to be limited by the resolutions of the virtual delay and virtual Doppler, respectively.

To overcome this challenge, \cite{WeiOGSBI} proposed  OGSBI. In OGSBI, on-grid and off-grid elements corresponding to indices of the virtual grid were introduced. Then,  $k_{p}$ and $l_{p}$ were estimated based on those on-grid and off-grid elements. Despite its benefits, when introducing the off-grid elements, OGSBI introduced a linear approximation error (see (22) in \cite{WeiOGSBI}), which brings about an approximation error. To overcome this limitation of OGSBI, the grid evolution process was introduced in a novel channel estimation scheme in \cite{Shan2} referred to as GESBI.
In GESBI, after the SBL iteration, the locations of the virtual DD grids  were adjusted to ensure the off-grid elements reach zero, which mitigates the approximation error to a certain extent. Despite the benefits of GESBI, its grid evolution process directly relies on the  delay and the Doppler estimates from the SBL, which limits the improvements of the channel estimation.

Inspired by these drawbacks of OGSBI and GESBI,  we propose   GRASBI  to adjust the virtual grid points and simultaneously update the channel parameters in the SBL. In particular, different from OGSBI and GESBI, our approach does not rely on any linear approximation when adjusting/refining the virtual grid points. Moreover,  the virtual grid refinement  and adjustment process does not rely on the delay and the Doppler estimated from the SBL. The details of the grid refinement and adjustment procedure are given next.

\textbf{\emph{Step 2: Likelihood-based Grid Points Refinement and Adjustment.}} In this step, the grid refinement and adjustment are performed. In particular,  as shown in Fig. $3$,  the grid refinement and adjustment are performed around the $\hat{P}$ top peaks of $\gammab$ one after the other. Denote the indices of the  $\hat{P}$ top peaks in $\gammab$ pseudospectrum as $\{i_{p},\ p\in\mathcal{I}_{\hat{P}}\}$.
Around every peak of $\gammab$, following two sub-steps are performed.

\textbf{\emph{Step 2.1: Grid Refinement.}} In this step,  we generate a refined virtual grid $\mathcal{A}_{p}$ around the $p$th peak. Specifically, as shown in Fig. $3$, the region $\mathcal{A}_{p}$ is divided into a uniform discrete grid with the size $\hat{M}\times\hat{N}$, where
$\hat{M}$ and $\hat{N}$ are the size along the delay and the Doppler dimensions, respectively. In doing so, we obtain $\mathcal{A}_{p}\triangleq\big\{(\bar{l}_{i},\bar{k}_{i})| \bar{l}_{i_{p}}\in[\bar{l}_{i_{p}}-\delta_{1},\bar{l}_{i_{p}}+\delta_{1}],\bar{k}_{i}\in[\bar{k}_{i_{p}}-\delta_{2},\bar{k}_{i_{p}}+\delta_{2}]\big\}$, where $\delta_{1}<r_{\tau}$ and $\delta_{2}<r_{\nu}$ to avoid grid overlap.

\textbf{\emph{Step 2.2:   Grid  Adjustment while Simultaneously Updating the Channel Estimation Parameters.}}  We utilize a ML-based scheme to adjust the grid points in the refined grid while updating the channel estimation parameters. The objective function of the ML principle is given as
\begin{subequations}
\begin{align}
\mathcal{L}(\Gammab)
=&\log p(\yb_{\text{T}}|\Gammab,\lambda)\\
=&\log\int_{-\infty}^{+\infty}P(\yb_{\text{T}}|\bar{\hb};\Gammab,\lambda)p(\bar{\hb}|\Gammab) {\rm d}\bar{\hb}\\
\triangleq& \log|\Cb|+\tr\big\{\Cb^{-1}\hat{\Rb}_{y}\big\}+\chi,
\end{align}
\end{subequations} where $\hat{\Rb}_{y}=\yb_{\text{T}}\yb^{\text{H}}_{\text{T}}$ is the sample covariance matrix, $\Cb\triangleq\lambda\Ib_{M_{\text{T}}N_{\text{T}}}+\bar{\Phib}\left(\bar{\lb},\bar{\kb}\right)\Gammab\bar{\Phib}^{\text{H}}\left(\bar{\lb},\bar{\kb}\right)$, and $\chi$ is a constant and can be ignored to simplify the expression of the objective function. To separate the $i_{p}$th grid component parameters from those of the rest of the other grid points, the objective function $\mathcal{L}(\Gammab)$ is rearranged, leading to the following lemma.
\begin{lemma}
The objective function of the ML principle can be expressed as
\begin{align}
\mathcal{L}(\Gammab)=&~ \mathcal{L}(\gamma_{i_{p}}, \bar{l}_{i_{p}},\bar{k}_{i_{p}})+\mathcal{L}(\gammab_{-i_{p}}),\label{eq:23}
\end{align}
where $\mathcal{L}(\gamma_{i_{p}}, \bar{l}_{i_{p}},\bar{k}_{i_{p}})$ denotes the term in the objective function corresponding to the $i_{p}$th element and $\mathcal{L}(\gammab_{-i_{p}})$ denotes the term in the objective function without the $i_{p}$th element. They can be expressed as
\begin{align}
\mathcal{L}(\gamma_{i_{p}},\bar{l}_{i_{p}},\bar{k}_{i_{p}})\triangleq&~ \log\left(1+\gamma_{i_{p}}\phib^{\text{H}}(\bar{l}_{i_{p}},\bar{k}_{i_{p}})\Cb^{-1}_{-i_{p}}\phib(\bar{l}_{i_{p}},\bar{k}_{i_{p}})\right)\nonumber\\
&\qquad-\frac{\phib^{\text{H}}(\bar{l}_{i_{p}},\bar{k}_{i_{p}})\Cb^{-1}_{-i_{p}}\hat{\Rb}_{y}\Cb^{-1}_{-i_{p}}\phib(\bar{l}_{i_{p}},\bar{k}_{i_{p}})}{\gamma^{-1}_{i_{p}}+\phib^{\text{H}}(\bar{l}_{i_{p}},\bar{k}_{i_{p}})\Cb^{-1}_{-i_{p}}\phib(\bar{l}_{i_{p}},\bar{k}_{i_{p}})},\label{eq:30}\\
\mathcal{L}(\gammab_{-i_{p}})\triangleq&~ \log\big|\Cb_{-i_{p}}\big|+\tr\left\{\Cb^{-1}_{-i_{p}}\hat{\Rb}_{y}\right\},\label{eq:31v3}\\
\Cb_{-i_{p}}\triangleq&~ \Phib_{-i_{p}}(\bar{\lb},\bar{\kb})\Gammab_{-i_{p}}\Phib_{-i_{p}}^{\text{H}}(\bar{\lb},\bar{\kb})+\lambda\Ib_{M_{\text{T}}N_{\text{T}}},
\end{align}
$\Phib_{-i_{p}}(\bar{\lb},\bar{\kb})$ denotes matrix $\Phib(\bar{\lb},\bar{\kb})$ without the $i_{p}$th column, $\Gammab_{-i_{p}}$ denotes matrix $\Gammab$ without the $i_{p}$th row and the $i_{p}$th column, and $\gammab_{-i_{p}}$ denotes vector $\gammab$ without the $i_{p}$th element.
\end{lemma}\quad\quad $\emph{Proof:}$ See Appendix A.$\hfill\square$

With the simplification in \eqref{eq:23}, we first update the $i_{p}$th grid point ($\bar{l}_{i_{p}},\bar{k}_{i_{p}}
$) by utilizing the grid points in the refinement grid. To this end, the term $\mathcal{L}(\gamma_{i_{p}}, \bar{l}_{i_{p}},\bar{k}_{i_{p}})$ is minimized while fixing $\mathcal{L}(\gammab_{-i_{p}})$ as
\begin{subequations}
\begin{align}
&\min_{(\bar{l}_{i_{p}},\bar{k}_{i_{p}})\in\mathcal{A}_{p}}\min_{\gamma_{i_{p}}\geq0} \mathcal{L}\left(\gamma_{i_{p}}, \bar{l}_{i_{p}},\bar{k}_{i_{p}}\right)\\
=&\min_{(\bar{l}_{i_{p}},\bar{k}_{i_{p}})\in\mathcal{A}_{p}}\min_{\gamma_{i_{p}}\geq0}\log\left(1+\gamma_{i_{p}} s(\bar{l}_{i_{p}},\bar{k}_{i_{p}})\right)-\frac{q(\bar{l}_{i_{p}},\bar{k}_{i_{p}})}{\gamma_{i_{p}}^{-1}+s(\bar{l}_{i_{p}},\bar{k}_{i_{p}})},
\end{align}
\end{subequations}
where
\begin{align}
q(\bar{l}_{i_{p}},\bar{k}_{i_{p}})\triangleq&~\phib^{\text{H}}(\bar{l}_{i_{p}},\bar{k}_{i_{p}})\Cb^{-1}_{-i_{p}}\hat{\Rb}_{y}\Cb^{-1}_{-i_{p}}\phib(\bar{l}_{i_{p}},\bar{k}_{i_{p}}), \label{eq:36}\\
s(\bar{l}_{i_{p}},\bar{k}_{i_{p}})\triangleq&~ \phib^{\text{H}}(\bar{l}_{i_{p}},\bar{k}_{i_{p}})\Cb^{-1}_{-i_{p}}\phib(\bar{l}_{i_{p}},\bar{k}_{i_{p}}).\label{eq:37}
\end{align}
Given  $\bar{k}_{i_{p}}$ and $\bar{l}_{i_{p}}$, the optimal value of $\gamma_{i_{p}}$  is obtained  as
\begin{subequations}
\begin{numcases}{\gamma^{\ast}_{i_{p}}=}
\frac{q(\bar{l}_{i_{p}},\bar{k}_{i_{p}})-s(\bar{l}_{i_{p}},\bar{k}_{i_{p}})}{s^{2}(\bar{l}_{i_{p}},\bar{k}_{i_{p}})},  \nonumber\\ {\qquad \qquad\qquad\quad q(\bar{l}_{i_{p}},\bar{k}_{i_{p}})>s(\bar{l}_{i_{p}},\bar{k}_{i_{p}}),} \label{eq:322a}    \\
0,      \qquad\qquad\qquad  {q(\bar{l}_{i_{p}},\bar{k}_{i_{p}})\leq s(\bar{l}_{i_{p}},\bar{k}_{i_{p}})}\label{eq:322b}.
\end{numcases}
\end{subequations}
Thereafter, substitute  $\gamma^{\ast}_{i_{p}}$  into $L(\gamma_{i},\bar{k}_{i_{p}}, \bar{l}_{i_{p}})$, we obtain
\begin{subequations}
\begin{numcases}{\mathcal{L}(\gamma^{\ast}_{i_{p}},\bar{l}_{i_{p}},\bar{k}_{i_{p}})=}
\log\frac{q(\bar{l}_{i_{p}},\bar{k}_{i_{p}})}{s(\bar{l}_{i},\bar{k}_{i})}-\frac{q(\bar{l}_{i},\bar{k}_{i_{p}})}{s(\bar{l}_{i},\bar{k}_{i})}+1, \nonumber \\ \qquad\  {q(\bar{l}_{i_{p}},\bar{k}_{i_{p}})>s(\bar{l}_{i_{p}},\bar{k}_{i_{p}}),}     \\
0,      \quad\ {q(\bar{l}_{i_{p}},\bar{k}_{i_{p}})\leq s(\bar{l}_{i_{p}},\bar{k}_{i_{p}})}.
\end{numcases}
\end{subequations}
It is noted that $\mathcal{L}(\gamma^{\ast}_{i_{p}},\bar{l}_{i_{p}},\bar{k}_{i_{p}})\leqslant0$ and is equal to zero only when $q(\bar{l}_{i_{p}},\bar{k}_{i_{p}})=s(\bar{l}_{i_{p}},\bar{k}_{i_{p}})$. In addition, $L(\gamma^{\ast}_{i_{p}},\bar{l}_{i_{p}},\bar{k}_{i_{p}})$ is a monotonically non-increasing function of $\frac{q(\bar{l}_{i_{p}},\bar{k}_{i_{p}})}{s(\bar{l}_{i_{p}},\bar{k}_{i_{p}})}$. Thus, the grid point adjustment is conducted to determine $\bar{l}_{i_{p}}$ and $\bar{k}_{i_{p}}$ based on the following maximization problem given by
\begin{align}
\left(\bar{l}^{\ast}_{i_{p}},\bar{k}^{\ast}_{i_{p}}\right)=\argmax_{(\bar{l}_{i_{p}},\bar{k}_{i_{p}})\in\mathcal{A}_{p},\ \atop \text{s.t.}\ q(\bar{l}_{i_{p}},\bar{k}_{i_{p}})>s(\bar{l}_{i_{p}},\bar{k}_{i_{p}}) } \frac{q(\bar{l}_{i_{p}},\bar{k}_{i_{p}})}{s(\bar{l}_{i_{p}},\bar{k}_{i_{p}})}.\label{eq:40}
\end{align}
We can see from (34) and (36) that the  grid points $\left(\bar{l}_{i_{p}},\bar{k}_{i_{p}}\right)$ and the parameters $\gamma_{i_{p}}$  in SBL
are adjusted simultaneously.
\begin{algorithm}[!t]
  \caption{: The Proposed GRASBI Algorithm}
  \begin{algorithmic}[1]
  \STATE{{\bfseries Input}:
  $\yb_{\text{T}}$, $\bar{\Phib}\left(\bar{\kb},\bar{\lb}\right)$, $M_{\text{T}}$, $N_{\text{T}}$, $M_{\tau}$, $N_{\nu}$, and $\Omegab_{0}$}; \\
  \STATE{ {\bf for} $\ell\in \mathcal{I}_{N_{\text{exter}}}$ }
  \STATE{\hspace{0.3cm}{\bf \underline{Step $1$: SBL with $\Omegab_{0}$:}}}
    \STATE{\quad Initialize $\lambda=\frac{\|\yb_{\text{T}}\|^{2}}{100M_{\text{T}}N_{\text{T}}}$ and $\Gammab=\Ib_{M_{\tau}N_{\nu}}$;}
    \STATE{\quad {\bf while} the variation of $\mub_{\bar{\hb}}$ is more than $\xi$  and the \hspace{0.8 cm} number  of
   the SBL iteration is less than $N_{\text{inter1}}$ { \bf  do}}
    \STATE{\qquad Step E:  Update $\Sigmab_{\bar{\hb}}$  using \eqref{eq:212} and $\mub_{\bar{\hb}}$  using \eqref{eq:222};}
    \STATE{\qquad Step M:  Update $\gamma_{i}$ using \eqref{eq:232} and $\lambda$ using \eqref{eq:242};}\\
  \STATE{\quad{\bfseries end while}}
  \STATE{\hspace{0.3cm}{\bf\underline{Step $2$:  Grid  refinement and adjustment:}}}
   \STATE{\quad {\bf while} the number of the grid adjustment is less than $N_{\text{inter2}}$ {\bf do}}
  \STATE{\quad{\bfseries for} $p\in\mathcal{I}_{\hat{P}}$ {\bfseries do}}
  \STATE{ \qquad Step $2.1$: Generate a refined virtual grid $\mathcal{A}_{p}$ around \\ \hspace{1.8cm} peak $p$;}
    \STATE{ \qquad Step $2.2$:  Given $\gammab_{-i_{p}}$ and $\Omegab_{p-1}$, update $\gamma_{i_{p}}$ and \\ \hspace{1.8 cm} $(\bar{l}_{i_{p}},\bar{k}_{i_{p}})$  using  (34) and (36), respectively;}
    \STATE{\quad {\bfseries end for}}
     \STATE{\quad{\bfseries end while}}
  \STATE{\quad$\Omegab_{0}=\Omegab_{\hat{P}}$}
  \STATE{{\bfseries end for}}
   \STATE{\bfseries Output}: $\mub_{\bar{\hb}}$, $\bar{\kb}$, and $\bar{\lb}$.
    \end{algorithmic}
\end{algorithm}

Consider the $\Omegab_{p}$ to be the virtual grid after updating the location of the $i_{p}$th peak in the virtual grid $\left(\bar{l}_{i_{p}},\bar{k}_{i_{p}}\right)$.
After updating  $\left(\bar{l}_{i_{p}},\bar{k}_{i_{p}}\right)$ and $\gamma_{i_{p}}$, as shown in Fig. $3$, the  update of $\left(\bar{l}_{i_{p+1}},\bar{k}_{i_{p+1}}\right)$ and the $\gamma_{i_{p}}$ corresponding to the $(p+1)$th peak are carried out based on $\Omegab_{p}$ and the $\gamma_{i_{p}}$.  This procedure is continued until the parameters including the $\gammab$ and the virtual grid points corresponding to all the $\hat{P}$ peaks are completed.

The summary of  GRASBI is provided in Algorithm $1$. The external iteration index is defined as $\ell$.
In the first external iteration ($\ell=1$), the SBL, with the maximum number of the internal iteration $N_{\text{inter1}}$,
is conducted based on the initial uniform grid, $\Omegab_{0}$ in $\textbf{Step 1}$. Then, the indices $i_{p}$, $p\in\mathcal{I}_{\hat{P}}$, are picked according to the peaks of the $\gammab$  pseudospectrum. Thereafter, the grid refinement and adjustment is carried out around each peak one after the other in $\textbf{Step 2}$. To this end, around each peak,
the refined virtual grid $\mathcal{A}_{p}$ is defined in $\textbf{Step 2.1}$. Then, the likelihood-based grid point adjustment is conducted to update the $\gamma_{i_{p}}$ while simultaneously adjusting the virtual grid point $(\bar{l}_{i_{p}},\bar{k}_{i_{p}})$ in $\textbf{Step 2.2}$.
After the $\hat{P}$ grid points are all updated $N_{\text{inter2}}$ times, the virtual grid is adjusted as $\Omegab_{\hat{P}}$.
Next, we update the virtual grid used in the next iteration as $\Omegab_{0}=\Omegab_{\hat{P}}$.
In the following external iterations ($\ell>1$), the SBL is rerun with the updated $\Omegab_{0}$.
Define  $N_{\text{exter}}$ as the maximum number of  the external  iteration.
The grid refinement and adjustment process continued until $N_{\text{exter}}$ is reached, i.e., $\ell=N_{\text{exter}}$, which can be decided based on complexity and estimation accuracy trade-off.

A visual illustration of our scheme for a channel with  three paths is presented in Fig. $3$.
We can see that the three paths are all off-grid in the original uniform virtual grid.
After the SBL in the first iteration, the three indices $i_{1}=10$, $i_{2}=30$, and $i_{3}=43$ are picked.
Then, the refined virtual grid $\mathcal{A}_{1}$ is defined around the $i_{1}$. The parameter $\gamma_{10}$ and the location of the grid point $(\bar{l}_{10}, \bar{k}_{10})$ are adjusted using (34) and (36), respectively, and the virtual DD grid $\Omegab_{0}$ is adjusted to $\Omegab_{1}$.
Thereafter, the refined virtual grid $\mathcal{A}_{2}$ is defined around the $i_{2}$.
With the $\Omegab_{1}$, the parameter $\gamma_{30}$  and  the $(\bar{l}_{30}, \bar{k}_{30})$ are adjusted, and the virtual grid $\Omegab_{1}$ is updated to the $\Omegab_{2}$.
Next, $\mathcal{A}_{3}$ is defined around the $i_{3}$.
The parameter $\gamma_{43}$  and the virtual grid point $(\bar{l}_{43}, \bar{k}_{43})$ are adjusted based on the $\Omegab_{2}$.
Finally,  the virtual grid is adjusted at $\Omegab_{3}$. In the next iteration, the SBL is performed while considering $\Omegab_{0}=\Omegab_{3}$.

\section{Low-complexity Channel Estimation Scheme}
In this section, the \textit{Student's t} distribution-based grid refinement and adjustment efficient sparse Bayesian inference scheme is proposed to reduce the computation complexity of the channel estimation scheme. Then, the complexity of the channel estimation schemes are analyzed.
\subsection{Low-complexity Grid Refinement and Adjustment-based  Channel Estimation Scheme}
Similar to GRASBI, T-GRAESBI involves two steps, namely efficient SBL and grid refinement and adjustment. However, different from GRASBI, the T-GRAESBI utilizes the efficient SBL to reduce the complexity of covariance matrix calculation as compared to \eqref{eq:212} in GRASBI.

In T-GRAEBI, the Student's t distribution is utilized to model the prior distribution rather than the Laplace distribution in GRASBI \cite{TIT}.
Moreover, the two-stage hierarchical prior is constructed as
\begin{align}
p(\bar{\hb}|\Gammab)=&~ \mathcal{CN}(\bar{\hb};\mathbf {0}, \Gammab^{-1}),  \label{eq:39V2}  \\
p(\Gammab)=&~ \prod_{i=0}^{M_{\tau}N_{\nu}-1}\Gamma(\gamma_{i};c_{0},d_{0}),\label{eq:40V2}
\end{align}
where $c_{0}, d_{0}>0$. The noise precision $\lambda^{-1}$ is assumed to follow the Gamma distribution as
\begin{align}
p(\lambda,c,d)=\Gamma(\lambda^{-1};c,d). \label{eq:41V2}
\end{align}

Based on \eqref{eq:39V2}-\eqref{eq:41V2} and utilizing the property of the continuously differentiable function, the exponential term $f(\bar{\hb})= \big\|\yb_{\text{T}}-\bar{\Phib}(\bar{\lb},\bar{\kb})\bar{\hb}\big \|^{2}$ of the likelihood function $P(\bar{\hb}|\yb_{\text{T}};\Gammab,\lambda)$ is approximated as
\begin{subequations}
\begin{align}
f(\bar{\hb})\leq&~ R(\bar{\hb},\xib)\\
            \triangleq&~ \|\yb_{\text{T}}-\bar{\Phib}(\bar{\lb},\bar{\kb})\xib \|^{2}+2(\bar{\hb}-\xib)^{\text{T}}\bar{\Phib}^{\text{T}}\nonumber\\
            &\qquad\times\left(\bar{\lb},\bar{\kb}\right)(\bar{\Phib}(\bar{\lb},\bar{\kb})\xib-\yb_{\text{T}})+s_{0}\|\bar{\hb}-\xib\|^{2},
\end{align}
\end{subequations}
where  $s_{0}=\text{eig}\left(\bar{\Phib}^{\text{H}}(\bar{\lb},\bar{\kb})\bar{\Phib}(\bar{\lb},\bar{\kb})\right)+\varepsilon$ is Lipschitz constant and $\varepsilon$ is a constant. Based on (40), the likelihood function can be expressed as the upper bound parameterized with $\xib$ as
\begin{subequations}
\begin{align}
p(\yb_{\text{T}}|\bar{\hb};\Gammab,\lambda)=&~ \max_{\xib}\hat{p}\left(\yb_{\text{T}}|\bar{\hb};\lambda,\xib \right)\\
=&~ \max_{\xib}\left(\frac{1}{2\pi\lambda}\right)^{\frac{M_{\text{T}}N_{\text{T}}}{2}}e^{-\frac{R(\bar{\hb},\xib)}{2\lambda}}.
\end{align}
\end{subequations}
The corresponding conditional posterior distribution in the $\imath$th iteration, $\imath\in\mathcal{I}_{N_{\text{inter1}}} $, can be expressed as
\begin{subequations}
\begin{align}
p(\bar{\hb}|\yb_{\text{T}};\Gammab[\imath],\lambda[\imath])\approx&~\frac{\hat{p}(\yb_{\text{T}}|\bar{\hb};\lambda[\imath],\xib[\imath])p(\bar{\hb};\Gammab[\imath])}{p(\yb_{\text{T}};\Gammab[\imath],\lambda[\imath])}\\
\propto&~ \mathcal{CN}(\hat{\mub}_{\bar{\hb}}[\imath],\hat{\Sigmab}_{\bar{\hb}}[\imath]),
\end{align}
\end{subequations}
where
\begin{align}
\hat{\Sigmab}_{\bar{\hb}}[\imath]=&~\left(\Gammab[\imath]+s_{0}\Ib_{M_{\tau}N_{\nu}} \right)^{-1},\label{eq:52}\\
\hat{\mub}_{\bar{\hb}}[\imath]=&~ (\lambda[\imath])^{-1}\hat{\Sigmab}_{\bar{\hb}}[\imath]\Big(s_{0}\xib[\imath]- \bar{\Phib}^{\text{H}}(\bar{\lb},\bar{\kb})\bar{\Phib}(\bar{\lb},\bar{\kb})\xib[\imath]\nonumber\\
&\qquad+ \bar{\Phib}^{\text{H}}(\bar{\lb},\bar{\kb})\yb_{\text{T}}\Big).\label{eq:53}
\end{align}
From \eqref{eq:52}, we can observe that the covariance matrix calculation of  T-GRAESBI  only involves a diagonal matrix inversion, and this reduces the complexity dramatically as compared to the covariance matrix determination of GRASBI in \eqref{eq:212}. With the above changes, the majorization-minimization framework is utilized to update the parameters. Based on these and the derivation results in \cite{ESBL}, and \cite{ShanTWC2}, we can obtain the updating rules for the hyper parameters as
\begin{align}
\xib[\imath+1]=&~ \hat{\mub}_{\bar{\hb}}[\imath+1],\label{eq:54}\\
\varpi[\imath+1]=&~ \sum_{i=0}^{M_{\tau}N_{\nu}-1}\frac{1}{\lambda[\imath]+s_{0}\gamma_{i}[\imath]},\label{eq:55}\\
\gamma_{i}[\imath+1]=&~ \sqrt{\frac{(c_{1}+1)s_{0}+c_{1}\lambda[\imath]\gamma_{i}[\imath]}{(\lambda[\imath]+s_{0}\gamma^{-1}_{i}[\imath])(2d_{0}+ \{\hat{\mub}_{\bar{\hb}}[\imath+1]\}^{2}_{i} )}},\label{eq:56}
\end{align}
where $c_{1}=2c_{0}-2$ and $n_{1}=M_{\tau}N_{\nu}+2-M_{\text{T}}N_{\text{T}}-2c$. The low-complexity T-GRAESBI scheme can be  summarized similar to that in Algorithm $1$, but with the differences being in line $6$ and line  $7$ in it. In particular, the  $\Sigmab_{\bar{\hb}}$ and $\mub_{\bar{\hb}}$ in line $6$ for T-GRAESBI  will be calculated based on \eqref{eq:52} and \eqref{eq:53}, and  \eqref{eq:232} and \eqref{eq:242} in line $7$ for T-GRAESBI will be calculated based on \eqref{eq:54},  \eqref{eq:56}, and \eqref{eq:57}, which is shown at the top of the next page.
\begin{table}[t!]
\setlength{\abovecaptionskip}{0pt}
\setlength{\belowcaptionskip}{0pt}\setlength{\abovecaptionskip}{0pt}
\setlength{\belowcaptionskip}{0pt}
\centering
\footnotesize
\newcommand{\tabincell}[2]{\begin{tabular}{@{}#1@{}}#2\end{tabular}}
\caption{The Number of Complex Multiplications Associated with Different Operations}
\begin{tabular}{lll}
\toprule  
Operation&Number of multiplications\\
\midrule  
\eqref{eq:212} &$\frac{2}{3}(M_{\text{T}}N_{\text{T}})^{3}+2M_{\text{T}}N_{\text{T}}M_{\tau}N_{\nu}\left(M_{\text{T}}N_{\text{T}}+1\right)$\\
\eqref{eq:222} &$\big[(M_{\tau}N_{\nu})^{2}+M_{\tau}N_{\nu}\big]M_{\text{T}}N_{\text{T}}$\\
\eqref{eq:232} &$3M_{\tau}N_{\nu}$\\
\eqref{eq:242} &$M_{\text{T}}N_{\text{T}}M_{\tau}N_{\nu}+M_{\text{T}}N_{\text{T}}+M_{\tau}N_{\nu}$\\
\eqref{eq:36} &$2\hat{P}[3(M_{\text{T}}N_{\text{T}})^{2}+2M_{\text{T}}N_{\text{T}}]$\\
\eqref{eq:40} &$2\hat{M}\hat{N}\hat{P}$\\
\eqref{eq:52} &$M_{\tau}N_{\nu}$\\
\eqref{eq:53} &$(3+M_{\text{T}}N_{\text{T}})(M_{\tau}N_{\nu})^{2}+M_{\tau}N_{\nu}(1+M_{\text{T}}N_{\text{T}})$\\
\eqref{eq:55} &$3M_{\tau}N_{\nu}$\\
\eqref{eq:56} &$7M_{\tau}N_{\nu}$\\
\eqref{eq:57} &$2M_{\text{T}}N_{\text{T}}$\\
\bottomrule 
\end{tabular}
\end{table}
\begin{table*}[t!]
\setlength{\abovecaptionskip}{0pt}
\setlength{\belowcaptionskip}{0pt}\setlength{\abovecaptionskip}{0pt}
\setlength{\belowcaptionskip}{0pt}
\centering
\footnotesize
\newcommand{\tabincell}[2]{\begin{tabular}{@{}#1@{}}#2\end{tabular}}
\caption{COMPUTATION COMPLEXITY OF DIFFERENT CHANNEL ESTIMATION SCHEMES}
\begin{tabular}{lll}
\toprule  
Operation&Number of multiplications\\
\midrule  
OGSBI \cite{WeiOGSBI} &$N_{\text{inter1}}\Big(\frac{2}{3}\left(M_{\text{T}}N_{\text{T}}\right)^{3}+2\left(M_{\text{T}}N_{\text{T}}\right)^{2}M_{\tau}N_{\nu}
+\left(M_{\tau}N_{\nu}\right)^{2}M_{\text{T}}N_{\text{T}}+4M_{\text{T}}N_{\text{T}}M_{\tau}N_{\nu}+4M_{\tau}N_{\nu}+M_{\text{T}}N_{\text{T}}+2\hat{P}(\hat{P}+1)\Big)$\\
GESBI \cite{Shan2} &$N_{\text{exter}}\bigg(2M_{\text{T}}N_{\text{T}}\hat{P}+N_{\text{inter1}}\Big(\frac{2}{3}\left(M_{\text{T}}N_{\text{T}}\right)^{3}+2\left(M_{\text{T}}N_{\text{T}}\right)^{2}M_{\tau}N_{\nu}
+\left(M_{\tau}N_{\nu}\right)^{2}M_{\text{T}}N_{\text{T}}+4M_{\text{T}}N_{\text{T}}M_{\tau}N_{\nu}$\\
&\qquad$+4M_{\tau}N_{\nu}+M_{\text{T}}N_{\text{T}}+2\hat{P}(\hat{P}+1)\Big)\bigg)$\\
T-GEESBI \cite{ShanTWC2}&$N_{\text{exter}}\bigg(2M_{\text{T}}N_{\text{T}}\hat{P}+ N_{\text{inter1}}\left(\left(3+M_{\text{T}}N_{\text{T}}\right)\left(M_{\tau}N_{\nu}\right)^{2}+M_{\tau}N_{\nu}\left(12+M_{\text{T}}N_{\text{T}}\right)+ 2M_{\text{T}}N_{\text{T}}+ 2\hat{P}(\hat{P}+1)    \right)\bigg)$\\
GRASBI &$N_{\text{exter}}\bigg(2\hat{P}N_{\text{inter2}}\Big(3 \left(M_{\text{T}}N_{\text{T}}\right)^{2}+2 M_{\text{T}}N_{\text{T}} +\hat{M}\hat{N}\Big)+N_{\text{inter1}}\Big(\frac{2}{3}\left(M_{\text{T}}N_{\text{T}}\right)^{3}
+2 \left(M_{\text{T}}N_{\text{T}}\right)^{2}M_{\tau}N_{\nu}+(M_{\tau}N_{\nu})^{2}M_{\text{T}}N_{\text{T}}$\\
&\qquad$+4 M_{\text{T}}N_{\text{T}}M_{\tau}N_{\nu}+4 M_{\tau}N_{\nu}+M_{\text{T}}N_{\text{T}}\Big) \bigg)$\\
T-GRAESBI &$N_{\text{exter}}\bigg(2\hat{P}N_{\text{inter2}}\Big(3 \left(M_{\text{T}}N_{\text{T}}\right)^{2}+2 M_{\text{T}}N_{\text{T}} +\hat{M}\hat{N}\Big)+N_{\text{inter1}}\bigg(\left(3+M_{\text{T}}N_{\text{T}}\right)\left(M_{\tau}N_{\nu}\right)^{2}+M_{\tau}N_{\nu}\left(12+M_{\text{T}}N_{\text{T}}\right)$\\
&\qquad$+ 2M_{\text{T}}N_{\text{T}}+ 2\hat{P}(\hat{P}+1)    \bigg) \bigg)$\\
\bottomrule 
\end{tabular}
\end{table*}

\newcounter{tempeqncnt2}
\setcounter{tempeqncnt2}{\value{equation}}
\setcounter{equation}{47}
\begin{figure*}[!t]
\normalsize
\begin{align}
\lambda[\imath+1]=&~ \frac{n_{1}+\sqrt{n^{2}_{1}+4 \varpi[\imath+1]\left( \big\|\yb_{\text{T}}-\bar{\Phib}^{\text{H}}(\bar{\lb},\bar{\kb})\bar{\Phib}(\bar{\lb},\bar{\kb})\xib[\imath+1] \big\|^{2}+2d  \right)    } }{2 \varpi[\imath+1]}\label{eq:57}
\end{align}
\setcounter{tempeqncnt2}{\value{equation}}
\hrulefill
\vspace*{3pt}
\end{figure*}

\subsection{Complexity Analyses}
In this subsection, the complexity of the channel estimation schemes are studied by evaluating the number of the complex multiplications associated with them. For the proposed GRASBI scheme, the computation of the SBL interior iteration involves the calculation of the covariance matrix in \eqref{eq:212}, the mean vector in \eqref{eq:222}, and the hyper parameters in \eqref{eq:232} and \eqref{eq:242}.
The operation in \eqref{eq:212} dictates computational complexity of the GRASBI, whereas  the operations including the \eqref{eq:52}, \eqref{eq:53}, \eqref{eq:55}, \eqref{eq:56}, and \eqref{eq:57},  dictate the computational complexity for T-GRAESBI. The number of complex multiplications associated with the above-mentioned operations are given in Table I.

The grid adjustment in \eqref{eq:40} and the parameter $\gamma_{i_{p}}$ update in (34) that are conducted in the exterior iteration, they  involve the calculation of the $q(\bar{l}_{i_{p}},\bar{k}_{i_{p}})$  and $s(\bar{l}_{i_{p}},\bar{k}_{i_{p}})$. The common operation in \eqref{eq:36} and \eqref{eq:37} is $\phib^{\text{H}}(\bar{l}_{i_{p}},\bar{k}_{i_{p}})\Cb^{-1}_{-i_{p}}$ and it involves $(M_{\text{T}}N_{\text{T}})^{2}$ complex multiplications. We can deduce that the calculation of the $\phib^{\text{H}}(\bar{l}_{i_{p}},\bar{k}_{i_{p}})\Cb^{-1}_{-i_{p}}\phib(\bar{l}_{i_{p}},\bar{k}_{i_{p}})$ adds an extra $M_{\text{T}}N_{\text{T}}$ complex multiplications. The calculation of the $\phib^{\text{H}}(\bar{l}_{i_{p}},\bar{k}_{i_{p}})\Cb^{-1}_{-i_{p}}\hat{\Rb}_{y}\Cb^{-1}_{-i_{p}}\phib(\bar{l}_{i_{p}},\bar{k}_{i_{p}})$ adds an extra $2(M_{\text{T}}N_{\text{T}})^{2}+M_{\text{T}}N_{\text{T}}$ complex multiplications. These lead to the total number of the complex multiplication in \eqref{eq:36}  is $2\hat{P}\big[3(M_{\text{T}}N_{\text{T}})^{2}+2M_{\text{T}}N_{\text{T}}\big]$. In addition, \eqref{eq:40} involves $2\hat{M}\hat{N}\hat{P}$ complex multiplications. The number of the complex multiplications  is also summarized in Table I.

Based on the above analyses, we can obtain that the total number of the proposed GRASBI scheme is $N_{\text{exter}}\Big(2\hat{P}\Big(3 \left(M_{\text{T}}N_{\text{T}}\right)^{2}+2 M_{\text{T}}N_{\text{T}} +\hat{M}\hat{N}\Big)+N_{\text{inter}}\Big(\frac{2}{3}\left(M_{\text{T}}N_{\text{T}}\right)^{3}
+2 \left(M_{\text{T}}N_{\text{T}}\right)^{2}M_{\tau}N_{\nu}+(M_{\tau}N_{\nu})^{2}M_{\text{T}}N_{\text{T}}+4 M_{\text{T}}N_{\text{T}}M_{\tau}N_{\nu}+4 M_{\tau}N_{\nu}+M_{\text{T}}N_{\text{T}}\Big) \Big)$, where $N_{\text{inter}}$ is the maximum number of the interior SBL iteration.
Moreover, the total number of the proposed T-GRAESBI  can be obtained as $N_{\text{exter}}\Big(2\hat{P}\Big(3 \left(M_{\text{T}}N_{\text{T}}\right)^{2}+2 M_{\text{T}}N_{\text{T}} +\hat{M}\hat{N}\Big)+N_{\text{inter}}\Big(\left(3+M_{\text{T}}N_{\text{T}}\right)\left(M_{\tau}N_{\nu}\right)^{2}+M_{\tau}N_{\nu}\left(12+M_{\text{T}}N_{\text{T}}\right)
+ 2M_{\text{T}}N_{\text{T}}+ 2\hat{P}(\hat{P}+1)    \Big) \Big)$. The results, along with the number of the complex multiplication associated with the other channel estimation schemes proposed in the literature, i.e., including OGSBI in \cite{WeiOGSBI},  GESBI in  \cite{Shan2}, and T-GEESBI in \cite{ShanTWC2}, are also summarized in Table II.

\section{Numerical Results}
In this section, we evaluate the accuracy and the efficiency of our proposed GRASBI and T-GRAESBI  schemes. The normalized mean square error (NMSE) of the equivalent sampled DD domain channel matrix, given by $\frac{\mid\Hb_{\text{DD}}-\hat{\Hb}_{\text{DD}}\mid^{2}}{\mid\Hb_{\text{DD}}\mid^{2}}$, is adopted as the metric to evaluate the accuracy of the schemes. Furthermore, the number of the complex multiplications is utilized to evaluate the complexity.
The  DD domain channel estimation schemes including  OGSBI in  \cite{WeiOGSBI}, GESBI in \cite{Shan2}, and T-GEESBI in \cite{ShanTWC2} are used as benchmarks.

\begin{figure*}[!t]
\begin{minipage}[t]{0.49\linewidth}
  \centering
  \includegraphics[width=3.3 in]{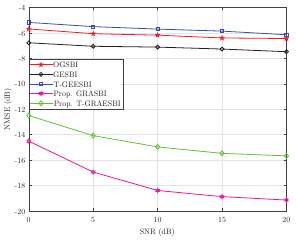}\\
 \caption{Illustration of the NMSE of  different channel estimation schemes. The results demonstrate that the proposed
GRASBI scheme outperforms all the other considered channel estimation schemes.}
  \end{minipage}
  \hfill
  \begin{minipage}[t]{0.49\linewidth}
  \centering
  \centerline{\includegraphics[width=3.3 in]{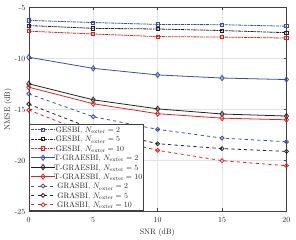}}
\caption{Illustration of the NMSE  for different channel estimation schemes  under different numbers of the exterior iterations $N_{\text{exter}}$. The results demonstrate that the accuracy of the channel estimation schemes improves with increasing the number of the exterior iterations.}
\end{minipage}%
\end{figure*}

Unless specified otherwise, we set the system parameter values as follows: $N=32$, $M=32$, the carrier frequency $f_{\text{c}}=4$ GHz,  $T=\frac{1}{15 \text{KHz}}$ , and the modulation scheme is QPSK. As in \cite{WeiOGSBI}, we set  the power of the pilot to be $10\log_{10}\frac{|d_{0}|^{2}}{|X_{\text{DD}}[l,k]|}=30$ $\mathrm{dB}$ higher than the information-bearing symbols.
For the square root Nyquist pulse used for pulse shaping, we set a square-root raised cosine pulse with the roll-off factor of $0.15$. For the channel, we set there are four channel taps and the power of each tap is uniformly distributed.  The Doppler of each tap is generated by the Jakes' formula $\nu_{p}=f_{\text{d}}\cos\theta_{p}$, $ p\in\mathcal{I}_{\hat{P}}$, where $f_{\text{d}}=\frac{v}{c}f_{\text{c}}$ and the angle of arrival of each tap is independently uniformly distributed in $(0, 2\pi]$. Considering the channel parameters, the maximum lag  and the maximum  normalized Doppler are set to $D=4$ and $k_{\text{max}}=4$, respectively. Based on these, we obtain  $M_{\text{T}}=D+1=5$ and $N_{\text{T}}=2k_{\text{max}}+1=9$.
With regards to the virtual DD grid, we set $M_{\tau}=N_{\nu}=10$.
For the parameters of the channel estimation schemes, we set $N_{\text{inter1}}=500$, $N_{\text{inter2}}=10$, $N_{\text{exter}}=2,5,10$, the threshold of  the iteration $\xi=10^{-3}$, the parameters of the prior distribution $c_{1}=2\times 10^{-6}$ and $d_{0}=10^{-6}$, the parameters of the noise precision $c=d=10^{-6}$, constant $\varepsilon=10^{-4}$.
The key findings are summarized as the observations.

\noindent
$\textbf{\emph{Observation 1:}}$
\noindent\emph{The proposed GRASBI outperforms the other DD domain channel estimation schemes. Specifically, GRASBI  has 11 dB NMSE performance gain at SNR=10 dB compared to OGSBI. (cf. Fig. $4$)}

In Fig. $4$, we compare the NMSE of the $\Hb_{\text{DD}}$ under $N_{\text{exter}}=5$ for the considered channel estimation schemes. We can see that our proposed GRASBI achieves the best performance among all the considered channel estimation schemes. In particular, GRASBI and T-GRAESBI  have $11$ $\mathrm{dB}$ and $8$ $\mathrm{dB}$ NMSE performance gains at $\mathsf{SNR}=10$ $\mathrm{dB}$ compared to OGSBI.
In addition, we observe that the proposed T-GRAESBI outperforms the grid evolution-based schemes GESBI and T-GEESBI, and the no grid evolution scheme OGSBI.
We highlight that these performance gains for our proposed GRASBI and T-GRAESBI come from the fact that while OGSBI, GESBI, and T-GEESBI consider linear approximation, no such approximation is considered in our proposed GRASBI and T-GRAESBI. We also clarify that the
difference in the updating rule for the virtual DD grid between
GRASBI and T-GRAESBI leads to their performance being different from one another.
It's interesting to find that T-GEESBI obtains the worst performance and this tendency is different from that in the OTFS system. Such inconsistency arises due to the different measurement matrices in the ODDM system  and the OTFS system.

\begin{figure*}[!t]
\begin{minipage}[t]{0.49\linewidth}
  \centering
  \includegraphics[width=3.3 in]{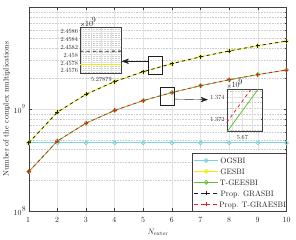}\\
 \caption{Illustration of the number of the complex multiplications under the different values of the $N_{\text{exter}}$. The results demonstrate that complexity of the proposed GRASBI is highest among the all the considered channel estimation schemes. In addition, the proposed T-GRAESBI reduces the complexity.}
  \end{minipage}
  \hfill
  \begin{minipage}[t]{0.49\linewidth}
  \centering
  \centerline{\includegraphics[width=3.3 in]{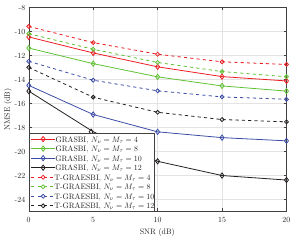}}
\caption{Illustration of the NMSE of different channel estimation schemes under the different  virtual DD grid sizes. The results show that the channel estimation accuracy improves with increasing the virtual DD grid size.}
\end{minipage}%
\end{figure*}

\noindent
$\textbf{\emph{Observation 2:}}$
\noindent\emph{The channel estimation accuracy improves when increasing the number of the $N_{\text{exter}}$. (cf. Fig. $5$)}

In Fig. $5$, we investigate the channel estimation accuracy  for our proposed GRASBI and T-GRAESBI under the different numbers of the exterior iterations, while considering GESBI as the benchmark.
We see that the channel estimation accuracy improves with increasing the value of  $N_{\text{exter}}$ for all the three channel estimation schemes.
For our proposed GRASBI and  T-GRAESBI, the estimation of the delay and the Doppler come from the virtual DD grids, which are updated according to the ML principle. When the  value of $N_{\text{exter}}$ increases, the virtual DD grids  get closer to the channel paths.
For GESBI, the estimation of the delay and the Doppler come from the sum of the on-grid and the off-grid elements. In particular, the higher value of $N_{\text{exter}}$ reduces the values of the off-grid elements, which mitigates the approximation error and improves the  accuracy.

\noindent
$\textbf{\emph{Observation 3:}}$
\noindent\emph{The  T-GRAESBI achieves a tradeoff between the accuracy and the complexity. (cf. Fig. $4$ and Fig. $6$)}

In Fig. $6$, we evaluate the complexity of the channel estimation schemes under the different numbers of the exterior iterations.
We first observe that with increasing the number of the exterior iterations, the number of the complex multiplications increases for all the  channel estimation schemes.
Second, we observe that the complexity of the proposed T-GRAESBI is lower than the proposed GRASBI.
This is because, with the utilization of the efficient SBL,  the proposed T-GRAESBI can significantly reduce the number of complex multiplications associated  the proposed GRASBI.
Third, we observe that the proposed GRASBI has a slightly higher complexity than GESBI due to the grid adjustment process.
Similar to that between  GESBI and  GRASBI,  T-GRAESBI also has a slightly higher complexity than T-GEESBI.
Based on the channel estimation accuracy comparison in Fig. $4$, we can arrive at the  conclusion that while the proposed GRASBI achieves the highest accuracy with the highest complexity,  the proposed T-GRAESBI provides a good tradeoff between the complexity  and the accuracy.

\begin{figure}[tbp]
\centering
\centerline{\includegraphics[width=3.3 in]{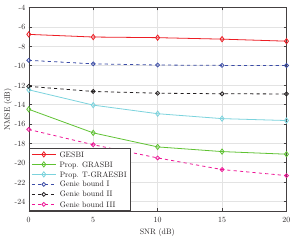}}
\caption{Comparison of the channel estimation accuracy of different channel estimation schemes with different genie bounds. The results show that the performance of the GRASBI outperform the genie bound I and genie bound II and have a smaller gap compared with the Genie bound III, which demonstrates the efficiency of the proposed scheme.}
\label{fig}
\end{figure}

\begin{figure*}[!t]
\begin{minipage}[t]{0.49\linewidth}
  \centering
  \includegraphics[width=3.3 in]{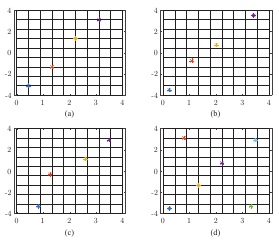}\\
 \caption{The uniform virtual DD grid under the four different channel conditions.}
  \end{minipage}
  \hfill
  \begin{minipage}[t]{0.49\linewidth}
  \centering
  \centerline{\includegraphics[width=3.3 in]{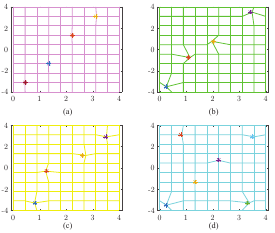}}
\caption{The results of the grid adjustment and refinement under the four conditions in Fig. $9$.}
\end{minipage}%
\end{figure*}

\noindent
$\textbf{\emph{Observation 4:}}$
\noindent\emph{The proposed low-complexity T-GRAESBI can outperform the proposed GRASBI when the virtual DD grid resolution of the former is higher than that of the later. Specifically,  T-GRAESBI with $M_{\tau}=N_{\nu}=10$ can outperform GRASBI with $M_{\tau}=N_{\nu}=8$. (cf. Fig. $7$)}

In Fig. $7$, we compare the channel estimation accuracy for the proposed GRASBI and T-GRAESBI under the different resolutions of the virtual DD grid and $N_{\text{exter}}=5$.
We first observe that under the fixed virtual grid size, the proposed GRASBI always outperforms the proposed T-GRAESBI.
Second, we observe that the channel estimation accuracy of both  GRASBI and T-GRAESBI improve with increasing the size of the virtual DD grid.
Larger the size of the virtual DD grid,  more sparse  the estimated channel vector and the more accurate the SBL estimation are. These provide more accurate initial point for the grid refinement and the adjustment,  which aligns with the results in Fig. $5$.
Third, we observe that when the virtual DD grid resolution of the proposed low-complexity T-GRAESBI is higher than that of the proposed GRASBI, the former can outperform the later.
Specifically,  T-GRAESBI under the conditions of   $M_{\tau}=N_{\nu}=10$ and
$M_{\tau}=N_{\nu}=12$ outperforms GRASBI under the conditions of $M_{\tau}=N_{\nu}=4$ and
$M_{\tau}=N_{\nu}=8$, respectively.
Fourth, we compare the number of the complex multiplications associated with the analyzed complexity results in Table II. Under
$M_{\tau}=N_{\nu}=10$ and the $M_{\tau}=N_{\nu}=12$,  the numbers of the complex multiplications for the T-GRAESBI are $1222723500\approx1.223\times 10^{9}$ and $2517313500\approx2.517\times 10^{9}$, respectively. The numbers of the complex multiplications for GRASBI under the conditions of  $M_{\tau}=N_{\nu}=4$ and  $M_{\tau}=N_{\nu}=8$ are $364011500\approx3.640\times 10^{8}$ and $1298892500\approx1.299\times 10^{9}>1.223\times 10^{9}$,  respectively.
Based on the third and fourth observations, we can conclude that through increasing the size of the virtual DD grid, T-GREASBI  can achieve better accuracy than  GRASBI while having a relatively low complexity.

\noindent
$\textbf{\emph{Observation 5:}}$
\noindent\emph{Both GRASBI and T-GRAESBI can outperform the channel estimation under the fixed and the uniform virtual DD grid conditions but are inferior to the channel estimation under the perfect grid adjustment condition. (cf. Fig. $8$)}

In Fig. $8$, we investigate  three types of the genie bounds associated with our proposed channel estimation schemes.
The genie bound I and the genie bound II are under the fixed and the uniform virtual DD grid. Specifically, the genie bound I is under the perfect on-grid elements information but ignores the off-grid elements. The genie bound II is under the perfect on-grid and the off-grid information but under the linear approximation system model.  The genie bound III is under the perfect grid adjustment condition which brings about no off-grid elements. We first observe that the performance of the genie bound II outperforms the genie bound I due to the consideration of the off-grid elements. Second, we observe that the genie bound III outperforms the genie bound I and genie bound I, which illustrate the significance of the virtual DD grid update to the channel estimation accuracy.
Third, we observe that the proposed GRASBI and T-GRAESBI outperform the genie bound II, but are inferior to the genie bound III. Finally, we observe that the performance of the GRASBI is  closer to the genie bound III.

\noindent
$\textbf{\emph{Observation 6:}}$
\noindent\emph{The proposed grid refinement and adjustment are robust under the different channel conditions. It can enable the distribution of the virtual DD grid near to the channel taps. (cf. Fig. $9$ and Fig. $10$)}

In Fig. $9$ and Fig. $10$, we compare the virtual DD grid before and after the grid refinement and adjustment for different channel conditions.
First, we  compare Fig. 9(a) with Fig. 10(a) for the scenario where all the channel taps have on-grid delays and on-grid Dopplers.
We observe that after the grid refinement and adjustment, the virtual DD grid remains nearly the same as before.
This demonstrates that the SBL algorithm in our proposed channel estimation schemes itself is sufficient to estimate the channel when the channel taps have only on-grid delays and on-grid Dopplers. In Fig. $9$(b), there are four channel taps with off-grid delays and off-grid Dopplers and the value of the off-grid parts are relatively large. We can see that after the grid refinement and adjustment, the virtual DD grid points move nearly to the channel paths position, which demonstrates the efficiency of the proposed off-grid channel estimation scheme. In Fig. $9$(c) there are four channel taps with off-grid delays and off-grid Dopplers and the off-grid elements are relatively small. We observe that similar to in Fig. $9$(a) and Fig. $9$(b), the virtual DD grids change to the position near to the channel paths. Finally, the Fig. $9$(d) demonstrates the condition of the six channel taps with each path is under the different condition. We can see that even under such a sophisticated condition, the grid refinement and adjustment can perform well to enable the virtual DD grid points near to the channel paths. Based on these observations, we can conclude that the proposed grid refinement and adjustment are robust.

\section{Conclusion}
In this paper, we propose two novel off-grid channel estimation schemes based on grid refinement and adjustment. First, the channel estimation problem was formulated as the sparse signal recovery through the definition of the virtual DD grid. Then,  GRASBI was proposed to improve the channel estimation accuracy by introducing the grid refinement and adjustment process after SBL. Specifically, the virtual grids were proposed to adjust according to the ML principle while simultaneously updating  the channel parameters. Next, the low-complexity T-GRAESBI scheme which utilizes the efficient SBL and the Student's t distribution was proposed. Moreover, the complexities of GRASBI and  T-GRAESBI are analyzed.
Finally, based on numerical results, we find that while the proposed GRASBI can provide the best channel estimation accuracy among all the channel estimation schemes for ODDM, the proposed T-GRAESBI provides a good tradeoff between the accuracy and the complexity.


\begin{appendices}
\section{Proof of Lemma $1$}
According to the definition of the $\Cb\triangleq \lambda\Ib_{M_{\text{T}}N_{\text{T}}}+\bar{\Phib}\left(\bar{\lb},\bar{\kb}\right)\Gammab\bar{\Phib}^{\text{H}}\left(\bar{\lb},\bar{\kb}\right)$, it can be rewritten as
\begin{subequations}
\begin{align}
\Cb
=&~ \lambda\Ib_{M_{\text{T}}N_{\text{T}}}+\sum_{m}\gamma_{m}\phib(\bar{l}_{m},\bar{k}_{m})\phib^{\text{H}}(\bar{l}_{m},\bar{k}_{m})\label{eq:41}\\
=&~ \lambda\Ib_{M_{\text{T}}N_{\text{T}}}+\sum_{m\neq i_{p}}\gamma_{m}\phib(\bar{l}_{m},\bar{k}_{m})\phib^{\text{H}}(\bar{l}_{m},\bar{k}_{m})\nonumber\\
&\qquad+\gamma_{i_{p}}\phib(\bar{l}_{i_{p}},\bar{k}_{i_{p}})\phib^{\text{H}}(\bar{l}_{i_{p}},\bar{k}_{i_{p}})\label{eq:42}\\
\triangleq&~ \Cb_{-i_{p}}+\gamma_{i_{p}}\phib(\bar{l}_{i_{p}},\bar{k}_{i_{p}})\phib^{\text{H}}(\bar{l}_{i_{p}},\bar{k}_{i_{p}}),\label{eq:43}
\end{align}
\end{subequations}
where \eqref{eq:41} follows from the matrix calculation; \eqref{eq:42} is obtained by diving the sum according to whether $m$ equals to the $i_{p}$ or not; and \eqref{eq:43} is obtained with the influence of the vector $\phib(\bar{l}_{i_{p}},\bar{k}_{i_{p}})$ removed.
Based on (49c), the inverse of the matrix $\Cb$ can be determined as
\begin{align}
|\Cb|=&~ \big|\Cb_{-i_{p}}\big|\Big|1+\gamma_{i_{p}}\phib^{\text{H}}(\bar{l}_{i_{p}},\bar{k}_{i_{p}})\Cb^{-1}_{-i_{p}}\phib(\bar{l}_{i_{p}},\bar{k}_{i_{p}})\Big|,\label{eq:44}\\
\Cb^{-1}=&~ \Cb^{-1}_{-i_{p}}-\frac{\Cb^{-1}_{-i_{p}}\phib(\bar{l}_{i_{p}},\bar{k}_{i_{p}})\phib^{\text{H}}(\bar{l}_{i_{p}},\bar{k}_{i_{p}})\Cb^{-1}_{-i_{p}}}{\gamma^{-1}_{i_{p}}+\phib^{\text{H}}(\bar{l}_{i_{p}},\bar{k}_{i_{p}})\Cb^{-1}_{-i_{p}}\phib(\bar{l}_{i_{p}},\bar{k}_{i_{p}})}.\label{eq:45}
\end{align}
Next, based on \eqref{eq:30}, \eqref{eq:31v3}, \eqref{eq:44}, and \eqref{eq:45}, the objective function can be further expressed as
\begin{subequations}
\begin{align}
\mathcal{L}(\Gammab)=&~ \log|\Cb_{-i_{p}}|+\log\left(1+\gamma_{i_{p}}\phib^{\text{H}}(\bar{l}_{i_{p}},\bar{k}_{i_{p}})\Cb^{-1}_{-i_{p}}\phib(\bar{l}_{i_{p}},\bar{k}_{i_{p}})\right)\nonumber\\
&\qquad-\tr\Bigg\{\frac{\Cb^{-1}_{-i_{p}}\phib(\bar{l}_{i_{p}},\bar{k}_{i_{p}})\phib^{\text{H}}(\bar{l}_{i_{p}},\bar{k}_{i_{p}})\Cb^{-1}_{-i_{p}}\hat{\Rb}_{y}}{\gamma^{-1}_{i_{p}}+\phib^{\text{H}}(\bar{l}_{i_{p}},\bar{k}_{i_{p}})\Cb^{-1}_{-i_{p}}\phib(\bar{l}_{i_{p}},\bar{k}_{i_{p}})}\Bigg\}\nonumber\\
&\qquad+\tr\Big\{\Cb^{-1}_{-i_{p}}\hat{\Rb}_{y}\Big\}\label{eq:46}\\
=&~ \log\left(1+\gamma_{i_{p}}\phib^{\text{H}}(\bar{l}_{i_{p}},\bar{k}_{i_{p}})\Cb^{-1}_{-i_{p}}\phib(\bar{l}_{i_{p}},\bar{k}_{i_{p}})\right)\nonumber\\
&\qquad-\frac{\phib^{\text{H}}(\bar{l}_{i_{p}},\bar{k}_{i_{p}})\Cb^{-1}_{-i_{p}}\hat{\Rb}_{y}\Cb^{-1}_{-i_{p}}\phib(\bar{l}_{i_{p}},\bar{k}_{i_{p}})}{\gamma^{-1}_{i_{p}}+\phib^{\text{H}}(\bar{l}_{i_{p}},\bar{k}_{i_{p}})\Cb^{-1}_{-i_{p}}\phib(\bar{l}_{i_{p}},\bar{k}_{i_{p}})}\\
&\qquad+\underbrace{\log|\Cb_{-i_{p}}|+\tr\Big\{\Cb^{-1}_{-i_{p}}\hat{\Rb}_{y}\Big\}}_{\mathcal{L}(\gammab_{-i_{p}})},\label{eq:47}
\end{align}
\end{subequations}
where \eqref{eq:46} is obtained by substituting \eqref{eq:44}  and \eqref{eq:45}  into (26c). Moreover, with the purpose of isolating the effect of the variable $\gamma_{i_{p}}$ and the grid point $\bar{l}_{i_{p}},\bar{k}_{i_{p}}$, \eqref{eq:47} is obtained by rearranging the order in \eqref{eq:46} and by utilizing the property of the matrix calculation.

\end{appendices}


\end{document}